\documentclass[twocolumn,twocolappendix]{aastex631}
\usepackage{hyperref}
\usepackage{amsmath}
\usepackage{mathtools}
\usepackage{soul}
\usepackage{longtable}
\usepackage{multirow}
\usepackage{bm}

\newcommand\swift{{\it Swift}}
\newcommand\thisgrb{GRB\,231117A}

\begin{document}

\title{The radio flare and multi-wavelength afterglow of the short \thisgrb: \\ energy injection from a violent shell collision}


\correspondingauthor{G. E. Anderson}
\email{gemma.anderson.astro@gmail.com}


\author[0000-0001-6544-8007]{G. E. Anderson}
\affiliation{International Centre for Radio Astronomy Research, Curtin University, GPO Box U1987, Perth, WA 6845, Australia}

\author[0000-0001-5169-4143]{G. P. Lamb}
\affiliation{Astrophysics Research Institute, Liverpool John Moores University, IC2 Liverpool Science Park, 146 Brownlow Hill, Liverpool, L3 5RF, UK}

\author[0000-0002-5826-0548]{B. P. Gompertz}
\affiliation{School of Physics and Astronomy, University of Birmingham, Birmingham B15 2TT, UK}
\affiliation{Institute for Gravitational Wave Astronomy, University of Birmingham, Birmingham B15 2TT}

\author[0000-0003-2705-4941]{L. Rhodes}
\affiliation{Astrophysics, Department of Physics, University of Oxford, Keble Road, Oxford, OX1 3RH, UK}
\affiliation{Trottier Space Institute at McGill, 3550 Rue University, Montreal, Quebec H3A 2A7, Canada}
\affiliation{Department of Physics, McGill University, 3600 Rue University, Montreal, Quebec H3A 2T8, Canada}

\author[0000-0001-5108-0627]{A. Martin-Carrillo}
\affiliation{School of Physics and Centre for Space Research, University College Dublin, Dublin, D04V1W8, Dublin, Ireland}

\author[0000-0001-9149-6707]{A. J. van der Horst}
\affiliation{Department of Physics, George Washington University, 725 21st St NW, Washington, DC 20052, USA}

\author[0000-0002-1195-7022]{A. Rowlinson}
\affiliation{Anton Pannekoek Institute for Astronomy, University of Amsterdam, Science Park 904, P.O. Box 94249, 1090GE Amsterdam, The Netherlands}
\affiliation{ASTRON, the Netherlands Institute for Radio Astronomy, Postbus 2, NL-7990 AA Dwingeloo, The Netherlands}


\author[0000-0003-1767-5277]{M. E. Bell}
\affiliation{Data Science Institute, University of Technology Sydney, 15 Broadway, Ultimo, NSW 2007, Australia}

\author[0000-0002-1066-6098]{T.-W. Chen}
\affiliation{Graduate Institute of Astronomy, National Central University, 300 Jhongda Road, 32001 Jhongli, Taiwan}

\author[0000-0002-2927-2398]{H. M. Fausey}
\affiliation{Department of Physics, George Washington University, 725 21st St NW, Washington, DC 20052, USA}

\author[0009-0007-5708-7978]{M. Ferro}
\affiliation{INAF - Osservatorio Astronomico di Brera, Via E. Bianchi 46, 23807 Merate (LC), Italy}

\author[0000-0002-4203-2946]{P. J. Hancock}
\affiliation{Curtin Institute for Data Science, Curtin University, GPO Box U1987, Perth, WA 6845, Australia}

\author[0000-0001-9309-7873]{S. R. Oates}
\affiliation{Department of Physics, Lancaster University, Lancaster, LA1 4YB, UK}

\author[0000-0001-6797-1889]{S. Schulze}
\affiliation{Center for Interdisciplinary Exploration and Research in Astrophysics (CIERA), Northwestern University, 1800 Sherman Ave., Evanston, IL 60201, USA}

\author[0000-0001-5108-0627]{R. L. C. Starling}
\affiliation{School of Physics and Astronomy, University of Leicester, University Road, Leicester LE1 7RH, UK}

\author[0000-0002-2898-6532]{S. Yang}
\affiliation{Henan Academy of Sciences, Zhengzhou 450046, Henan, China}


\author[0000-0002-8648-0767]{K. Ackley}
\affiliation{Department of Physics, University of Warwick, Coventry, CV4 7AL, UK}

\author[0000-0003-0227-3451]{J. P. Anderson}
\affiliation{European Southern Observatory, Alonso de C\'ordova 3107, Casilla 19, Santiago, Chile}
\affiliation{Millennium Institute of Astrophysics MAS, Nuncio Monsenor Sotero Sanz 100, Off.
104, Providencia, Santiago, Chile}

\author[0000-0003-2734-1895]{A. Andersson}
\affiliation{Astrophysics, Department of Physics, University of Oxford, Keble Road, Oxford, OX1 3RH, UK}

\author[0000-0001-6991-7616]{J. F. Agüí Fernández}
\affiliation{Centro Astron\'omico Hispano en Andaluc\'ia, Observatorio de Calar Alto, Sierra de los Filabres, G\'ergal, Almer\'ia, 04550, Spain}

\author[0009-0000-0564-7733]{R. Brivio}
\affiliation{INAF - Osservatorio Astronomico di Brera, Via E. Bianchi 46, 23807 Merate (LC), Italy}

\author[0000-0002-2942-3379]{E. Burns}
\affiliation{Department of Physics \& Astronomy, Louisiana State University, Baton Rouge, LA 70803, USA}

\author[0000-0001-6965-7789]{K. C. Chambers}
\affiliation{Institute for Astronomy, University of Hawaii, 2680 Woodlawn Drive, Honolulu HI 96822, USA}

\author[0000-0001-5486-2747]{T. de Boer}
\affiliation{Institute for Astronomy, University of Hawaii, 2680 Woodlawn Drive, Honolulu HI 96822, USA}

\author[0000-0002-7320-5862]{V. D'Elia}
\affiliation{Space Science Data Center (SSDC) - Agenzia Spaziale Italiana (ASI), Via del Politecnico snc, 00133 Roma, Italy}

\author[0000-0002-4036-7419]{M. De Pasquale}
\affiliation{Department of Mathematics and Computer Sciences, Physical Sciences and Earth Sciences, University of Messina, Via F.S. D’Alcontres 31, Messina, 98166, Italy}

\author[0000-0001-7717-5085]{A. de Ugarte Postigo}
\affiliation{Aix Marseille University, CNRS, CNES, LAM, Marseille, France}

\author[0000-0001-9868-9042]{Dimple}
\affiliation{School of Physics and Astronomy, University of Birmingham, Birmingham B15 2TT, UK}
\affiliation{Institute for Gravitational Wave Astronomy, University of Birmingham, Birmingham B15 2TT}

\author[0000-0002-5654-2744]{R. Fender}
\affiliation{Astrophysics, Department of Physics, University of Oxford, Keble Road, Oxford, OX1 3RH, UK}

\author[0000-0003-1916-0664]{M. D. Fulton}
\affiliation{Astrophysics Research Centre, School of Mathematics and Physics, Queen’s University Belfast, BT7 1NN, UK}

\author[0000-0003-1015-5367]{H. Gao}
\affiliation{Institute for Astronomy, University of Hawaii, 2680 Woodlawn Drive, Honolulu HI 96822, USA}

\author[0000-0002-8094-6108]{J. H. Gillanders}
\affiliation{Astrophysics, Department of Physics, University of Oxford, Keble Road, Oxford, OX1 3RH, UK}

\author[0000-0003-3189-9998]{D. A. Green}
\affiliation{Astrophysics Group, Cavendish Laboratory, 19 J.J. Thomson Avenue, Cambridge CB3 0HE, UK}

\author[0000-0002-1650-1518]{M. Gromadzki}
\affiliation{Astronomical Observatory, University of Warsaw, Al. Ujazdowskie 4,
00-478 Warszawa, Poland}

\author[0000-0001-8802-520X]{A. Gulati}
\affiliation{Sydney Institute for Astronomy, School of Physics, The University of Sydney, NSW 2006, Australia}
\affiliation{ARC Centre of Excellence for Gravitational Wave Discovery (OzGrav), Hawthorn, VIC 3122, Australia}
\affiliation{CSIRO Space and Astronomy, PO Box 76, Epping, NSW 1710, Australia}

\author[0000-0002-8028-0991]{D. H. Hartmann}
\affiliation{Department of Physics and Astronomy, Clemson University, Clemson, SC 29634-0978, USA}

\author[0000-0003-1059-9603]{M. E. Huber}
\affiliation{Institute for Astronomy, University of Hawaii, 2680 Woodlawn Drive, Honolulu HI 96822, USA}

\author[0000-0003-4650-4186]{N. P. M. Kuin}
\affiliation{University College London, Mullard Space Science Laboratory, Holmbury St. Mary, Dorking, RH5 6NT, UK}

\author[0000-0002-9415-3766]{J. K.\ Leung}
\affiliation{David A. Dunlap Department of Astronomy and Astrophysics, University of Toronto, 50 St. George Street, Toronto, ON M5S 3H4, Canada}
\affiliation{Dunlap Institute for Astronomy and Astrophysics, University of Toronto, 50 St. George Street, Toronto, ON M5S 3H4, Canada}
\affiliation{Racah Institute of Physics, The Hebrew University of Jerusalem, Jerusalem 91904, Israel}

\author[0000-0001-7821-9369]{A. J. Levan}
\affiliation{Department of Astrophysics/IMAPP, Radboud University Nijmegen, P.O. Box 9010, Nijmegen, 6500 GL, The Netherlands}

\author[0000-0002-7272-5129]{C.-C. Lin}
\affiliation{Institute for Astronomy, University of Hawaii, 2680 Woodlawn Drive, Honolulu HI 96822, USA}

\author[0000-0002-7965-2815]{E. Magnier}
\affiliation{Institute for Astronomy, University of Hawaii, 2680 Woodlawn Drive, Honolulu HI 96822, USA}

\author[0000-0002-7517-326X]{D. B. Malesani}
\affiliation{Cosmic Dawn Center (DAWN), Denmark}
\affiliation{Niels Bohr Institute, University of Copenhagen, Jagtvej 128, Copenhagen, 2200, Denmark}
\affiliation{Department of Astrophysics/IMAPP, Radboud University Nijmegen, P.O. Box 9010, Nijmegen, 6500 GL, The Netherlands}

\author[0009-0003-8803-8643]{P. Minguez}
\affiliation{Institute for Astronomy, University of Hawaii, 2680 Woodlawn Drive, Honolulu HI 96822, USA}

\author[0000-0002-2557-5180]{K. P. Mooley}
\affiliation{Indian Institute Of Technology Kanpur, Kanpur, Uttar Pradesh 208016, India}
\affiliation{Caltech, 1200 E. California Blvd.  MC 249-17, Pasadena, CA 91125, USA}

\author[0009-0004-7639-869X]{T. Mukherjee}
\affiliation{School of Mathematical and Physical Sciences, Macquarie University, Sydney, NSW 2109, Australia}
\affiliation{Astrophysics and Space Technologies Research Centre, Macquarie University, Sydney, NSW 2109, Australia}

\author[0000-0002-2555-3192]{M. Nicholl}
\affiliation{Astrophysics Research Centre, School of Mathematics and Physics, Queens University Belfast, Belfast BT7 1NN, UK}

\author[0000-0002-5128-1899]{P. T. O'Brien}
\affiliation{School of Physics and Astronomy, University of Leicester, University Road, Leicester LE1 7RH, UK}

\author[0000-0003-3457-9375]{G. Pugliese}
\affiliation{Anton Pannekoek Institute for Astronomy, University of Amsterdam, Science Park 904, P.O. Box 94249, 1090GE Amsterdam, The Netherlands}

\author[0000-0002-8860-6538]{A. Rossi}
\affiliation{INAF - Osservatorio di Astrofisica e Scienza dello Spazio, via Piero Gobetti 93/3, 40129 Bologna, Italy}

\author[0000-0003-4501-8100]{S. D. Ryder}
\affiliation{School of Mathematical and Physical Sciences, Macquarie University, Sydney, NSW 2109, Australia}
\affiliation{Astrophysics and Space Technologies Research Centre, Macquarie University, Sydney, NSW 2109, Australia}

\author[0000-0001-6620-8347]{B. Sbarufatti}
\affiliation{INAF - Osservatorio Astronomico di Brera, Via E. Bianchi 46, 23807 Merate (LC), Italy}

\author[0000-0003-4876-7756]{B. Schneider}
\affiliation{Aix Marseille University, CNRS, CNES, LAM, Marseille, France}
\affiliation{Massachusetts Institute of Technology, Kavli Institute for Astrophysics and Space Research, Cambridge, Massachusetts, USA}

\author[0000-0003-1500-6571]{F. Sch\"ussler}
\affiliation{IRFU, CEA, Université Paris-Saclay, F-91191 Gif-sur-Yvette, France}

\author[0000-0002-8229-1731]{S. J. Smartt}
\affiliation{Astrophysics, Department of Physics, University of Oxford, Keble Road, Oxford, OX1 3RH, UK}
\affiliation{Astrophysics Research Centre, School of Mathematics and Physics, Queen’s University Belfast, BT7 1NN, UK}

\author[0000-0001-9535-3199]{K. W. Smith}
\affiliation{Astrophysics, Department of Physics, University of Oxford, Keble Road, Oxford, OX1 3RH, UK}
\affiliation{Astrophysics Research Centre, School of Mathematics and Physics, Queen’s University Belfast, BT7 1NN, UK}

\author[0000-0003-4524-6883]{S. Srivastav}
\affiliation{Astrophysics, Department of Physics, University of Oxford, Keble Road, Oxford, OX1 3RH, UK}

\author[0000-0003-0771-4746]{D. Steeghs}
\affiliation{Department of Physics, University of Warwick, Gibbet Hill Road, Coventry CV4 7AL, UK}

\author[0000-0003-3274-6336]{N. R. Tanvir}
\affiliation{School of Physics and Astronomy, University of Leicester, University Road, Leicester LE1 7RH, UK}

\author[0000-0002-7978-7648]{C. C. Thoene}
\affiliation{E. Kharadze Georgian National Astrophysical Observatory, Mt. Kanobili, Abastumani 0301, Adigeni, Georgia}

\author[0000-0001-9398-4907]{S. D. Vergani}
\affiliation{LUX, Observatoire de Paris, Université PSL, CNRS, Sorbonne Université, 92190 Meudon, France}

\author[0000-0002-1341-0952]{R. J. Wainscoat}
\affiliation{Institute for Astronomy, University of Hawaii, 2680 Woodlawn Drive, Honolulu HI 96822, USA}

\author{Z.-N. Wang}
\affiliation{Henan Academy of Sciences, Zhengzhou 450046, Henan, China}

\author[0000-0002-3101-1808]{R. A. M. J. Wijers}
\affiliation{Anton Pannekoek Institute for Astronomy, University of Amsterdam, Science Park 904, P.O. Box 94249, 1090GE Amsterdam, The Netherlands}

\author[0000-0001-7361-0246]{D. Williams-Baldwin}
\affiliation{Jodrell Bank Centre for Astrophysics, School of Physics and Astronomy, The University of Manchester, Manchester, M13 9PL, UK}

\author[0009-0007-3476-2272]{I. Worssam}
\affiliation{School of Physics and Astronomy, University of Birmingham, Birmingham B15 2TT, UK}
\affiliation{Institute for Gravitational Wave Astronomy, University of Birmingham, Birmingham B15 2TT}

\author[0000-0003-3935-7018]{T. Zafar}
\affiliation{School of Mathematical and Physical Sciences, Macquarie University, Sydney, NSW 2109, Australia}



\begin{abstract}

We present the early radio detection and multi-wavelength modeling of the short gamma-ray burst (GRB) 231117A at redshift $z=0.257$. The Australia Telescope Compact Array automatically triggered a 9-hour observation of \thisgrb\ at 5.5 and 9\,GHz following its detection by the \textit{Neil Gehrels Swift Observatory} just 1.3\,hours post-burst. 
Splitting this observation into 1-hour time bins, the early radio afterglow exhibited flaring, scintillating and plateau phases. 
The scintillation allowed us to place the earliest upper limit ($<10$\,hours) on the size of a GRB blast wave to date, constraining it to  $<1\times10^{16}$\,cm. 
Multi-wavelength modeling of the full afterglow required a period of significant energy injection between $\sim 0.02$ and $1$\,day. 
The energy injection was modeled as a violent collision of two shells: a reverse shock passing through the injection shell explains the early radio plateau, while an X-ray flare is consistent with a shock passing through the leading impulsive shell.
Beyond 1\,day, the blast wave evolves as a classic decelerating forward shock with an electron distribution index of $p=1.66\pm0.01$.
Our model also indicates a jet-break at $\sim2$\,days, and a half-opening angle of $\theta_j=16\mathring{.}6 \pm 1\mathring{.}1$.
Following the period of injection, the total energy is $\zeta\sim18$ times the initial impulsive energy, with a final collimation-corrected energy of $E_{\mathrm{Kf}}\sim5.7\times10^{49}$\,erg.
The minimum Lorentz factors this model requires are consistent with constraints from the early radio measurements of $\Gamma>35$ to $\Gamma>5$ between $\sim0.1$ and $1$\,day.
These results demonstrate the importance of rapid and sensitive radio follow-up of GRBs for exploring their central engines and outflow behaviour.

\end{abstract}

\keywords{}

\section{Introduction} \label{sec:intro}

Short gamma-ray bursts (SGRBs) are extremely high-energy transient events with durations of 2 seconds or less. They are thought to be driven by the mergers of binary neutron star (BNS) systems, a scenario that received significant support following the first gravitational-wave (GW) and electromagnetic (EM) multi-messenger detection \citep{Abbott17a} of GW170817 \citep{Abbott17b} and GRB 170817A \citep{Goldstein17,Savchenko17}. 
Canonically, SGRBs are separated from long GRBs (LGRBs) based on an observed bimodality in duration (measured as $t_{90}$; the time in which the middle 90 per cent of the gamma-ray fluence is emitted) and spectral hardness \citep{Kouveliotou93}. 
The divide of $t_{90} \leq 2$\,s for SGRBs and $t_{90} > 2$\,s for LGRBs ostensibly maps to their disparate progenitors. For SGRBs, the 2\,s limit roughly corresponds to the maximum time in which the remnant black hole formed in the merger is expected to accrete the residual torus, which powers the jet \citep[e.g.][]{Ruffert99,Rosswog03,Hotokezaka11}. Conversely, since LGRBs are driven by the core-collapse of very massive stars \citep{Galama98,Hjorth03,Stanek03}, their accretion discs are much more massive, allowing a far longer accretion timescale.

Several observational factors demonstrate that some BNS mergers could have periods of energy injection and/or prolonged central engine activity. 
When considering their prompt gamma-ray emission, $\sim25\%$ of SGRBs display a rebrightening in the gamma-rays following the initial emission spike known as extended emission \citep{lazzati01,norris06,Norris10}.
Explanations include magnetically driven winds from a long-lived magnetar engine, r-process heating on fallback accretion, and a two-jet model \citep{metzger08,metzger10,barkov11,bucciantini12,Gompertz13}.
Meanwhile, there have been two GRBs with $t_{90} > 2$\,s, where optical spectroscopy revealed an associated kilonova, powered by the decay of radioactive elements and likely created from compact binary mergers \citep[GRB 211211A and GRB 230307A;][]{rastinejad22,troja22nat,yang22nat,levan24,yang24nat}.
These results suggest that such systems can drive longer duration prompt emission.

Evidence for unusual or prolonged central engine activity from SGRBs may also be imprinted on their afterglows. 
Following the launch of a jet, the fireball model \citep{rees92,kobayashi97,sari97,piran99,kobayashi00sari} forecasts synchrotron emission components from a forward and reverse shock generated by relativistic ejecta interacting with the circumburst medium (CBM). 
However, often the multi-wavelength afterglows from both long and short GRBs are observed to plateau between minutes to hours post-burst \citep{zhang06}, with X-ray plateaus observed in $\sim50$\% of SGRB X-ray afterglows \citep{rowlinson13} plus numerous in the optical
\citep[e.g.][]{roming06,stratta07,fan13,deugartepostigo14,knust17,dainotti20,fernandez23}. 
Plateaus have also been detected in LGRB radio light curves \citep{levine22}.
Such plateaus suggest periods of energy injection beyond what is expected from accretion onto a black hole.
Indeed, there exists an intrinsic correlation between the plateau's rest frame end time and its corresponding luminosity \citep{dainotti08} as shown in the X-ray, optical, high-energy gamma-ray and radio bands \citep{dainotti13,dainotti20,dainotti21,levine22}. This correlation holds irrespective of the GRB classification, which suggests plateaus have a fixed energy reservoir \citep{dainotti20}.

The origin of the plateau emission is typically explained as the collision between slower shells in the jetted outflow catching a leading fast shell as it decelerates.
The energy is then injected in the form of mass, resulting in afterglow emission that declines more slowly than expected or undergoes a flaring episode \citep{meszaros97, rees98, kumar2000, zhang2002}.
Alternatively, energy could be injected by
magnetically dominated outflows from a long-lived magnetar central engine \citep{zhang2002, zhang06, metzger11,rowlinson13,Gompertz13,rowlinson14}, 
-- in such cases, the energy injection is always `mild' (i.e., no flares).
Where the sudden collision of massive shells causes injection, a shock system can be established that results in flaring \citep{ioka2005}, which can also explain variability within GRB afterglows. 
We therefore require extensive multi-wavelength investigations to understand the source of the plateau and flaring emission, which will enable us to explore the apparent diversity in SGRB central engine activity times and outflow behaviours.

Radio observations are necessary for breaking the degeneracies between the blast wave energetics and the density of the CBM that exists when only modeling the X-ray and optical afterglow \citep[e.g.][]{rastinejad22}. They also reveal unexpected features in both long and short GRB radio afterglows such as early \citep{laskar18,laskar19,anderson23} and late time flares \citep{schroeder24}, or additional synchrotron emitting components potentially from a cocoon or wider outflow \citep{rhodes22,rhodes24,leung25pp}. 
In particular, \citet{schroeder24} detected a radio flare from short GRB 210726A that began 11\,days post-burst that indicated an increase in the isotropic kinetic energy by a factor of four. Such a flare might have been caused by energy injection or from a reverse shock caused by a shell collision that would have required extreme late-time central engine activity.  
Very early-time radio follow-up of short GRB 231117A also identified the addition of a reverse shock that occurred $<1$\,day post-burst, either from the GRB jet interacting with the CBM or from an energy injection episode \citep{schroeder25}. 
This demonstrates that radio observations can provide vital 
insight into energy injection and/or prolonged central engine activity. 

Here we report an independent study of the \thisgrb\ afterglow using new radio and optical data, which has been supplemented by the \citet{schroeder25} dataset.  
The new radio and optical observations are outlined in Section~\ref{sec:obs_results}. Section~\ref{sec:wave} describes the wavelength-dependent analysis of \thisgrb, while Section~\ref{sec:afterglow_model} focuses on the multi-wavelength modeling of the afterglow.
In Section~\ref{sec:discussion}, we discuss how the radio properties of \thisgrb\ compare to the radio-detected SGRB population. This is followed by a discussion on energy injection, energetics and the efficiency of \thisgrb\ and how it compares to other SGRBs. We finish our discussions by interpreting the origin of the early radio emission from \thisgrb.
Throughout this paper, we have used the following power law representation for flux density, $S_{\nu} \propto t^{-\alpha}\nu^{-\beta}$, and assumed a cosmology of $H_0=67.4$ and $\Omega_m=0.315$ \citep{planck20}

\section{Observations and Results} \label{sec:obs_results}

\thisgrb\ was detected by the \emph{Swift} Burst Alert Telescope \citep[BAT;][]{Barthelmy05} at 03:03:19 UT on November 17$^{\rm th}$ 2023 with a peak count rate of $\sim 80,000$\,cts\,s$^{-1}$ \citep{laha23gcn}. It had a gamma-ray fluence of $2.3\pm0.1 \times 10^{-6}$\,erg\,cm$^{-2}$ (15--150\,keV band) and a $t_{90}$ (time range over which the GRB emits 5\% to 95\% of its counts) of $0.67\pm0.07$\,s, placing it firmly within the SGRB class \citep{markwardt23}. 
Detections of the prompt emission were also reported by AstroSat, the Astro-Rivelatore Gamma a Immagini Leggero (AGILE), Konus-Wind, Glowbug, GRBAlpha, the Gravitational Wave High-energy Electromagnetic Counterpart All-sky Monitor (GECAM) C telescope and the Calorimetric Electron Telescope  \citep{navaneeth23,cattaneo23,svinkin23,cheung23,dafcikova23,xue23,yamaoka23}. 

The GRB was quickly localised by the \emph{Swift} X-ray Telescope \citep[XRT;][]{Burrows05} through the detection of its X-ray afterglow \citep{beardmore23,melandri23}. 
The candidate optical afterglow and host galaxy were first reported by the 40\,cm SLT telescope at Lulin Observatory in an observation taken at 10:26 UT on November 17, 2023 (MJD = 60265.435), 7.39 hours post-burst \citep{yang23GCN}. The reported position was
RA (J2000): $22^{\rm h} 09^{\rm m} 33\fs37$, Dec (J2000): $+13\degr 31' 20\farcs2$, approximately 3 arcsec away from the enhanced Swift-XRT location \citep[2.0 arcsec radius, 90\% confidence;][]{beardmore23}, which was confirmed by observations with Keck \citep{rastinejad23}. 
The spectroscopic redshift of $z=0.257$ was obtained with Keck \citep{ahumada2023} and confirmed by \citet{gonzalex-ba23GCN} via the advanced Public ESO Spectroscopic Survey for Transient Objects \citep[ePESSTO+;][]{smartt15}.
The radio afterglow was quickly identified using the Australia Telescope Compact Array (ATCA) by \citet{rhodes23gcn_atca} and confirmed with observations from the Karl G. Jansky Very Large Array (VLA) by \citet{schroeder23GCN}. The afterglow was comprehensively observed in the radio, optical and X-ray bands up to nearly 50\,days post-burst.

In the following sections, we present the observations and results associated with the radio, optical and X-ray data analysed for the modeling in this paper, providing a comprehensive dataset with dense sampling in all bands from minutes to days post-burst.

\subsection{Radio Observations}\label{sec:radio}

\subsubsection{ATCA}\label{sec:radio_atca}

ATCA is a six-element interferometer of 22\,m diameter dishes based near Narrabri, Australia. 
Following the \swift-BAT notification of \thisgrb, the ATCA rapid-response mode \citep{anderson21} automatically triggered observations under programme C3204 (PI: Anderson), resulting in one of the earliest radio detections of a GRB afterglow \citep{rhodes23gcn_atca}.
The ATCA rapid-response observation began at 04:20\,UT, 1.3\,h post-burst as soon as the array had finished maintenance. The array observed for 9\,h using the 4\,cm dual receiver (centred at 5.5 and 9\,GHz, each with 2\,GHz bandwidth). Further ATCA follow-up observations were manually scheduled and conducted at 1.3, 3.1, 9.1 and 16.1\,days post-burst, in each case using the 4\,cm receiver and with some dates also including the 15\,mm receiver (central frequencies of 16.7 and 21.2\,GHz, note that poor weather resulted in the 21.2\,GHz observations being unusable). All data was processed using {\sc Miriad} \citep{sault95} using PKS 1934$-$638 and PKS 2230+114 as the flux and phase calibrator, respectively.

\begin{figure*}
    \centering
    \includegraphics{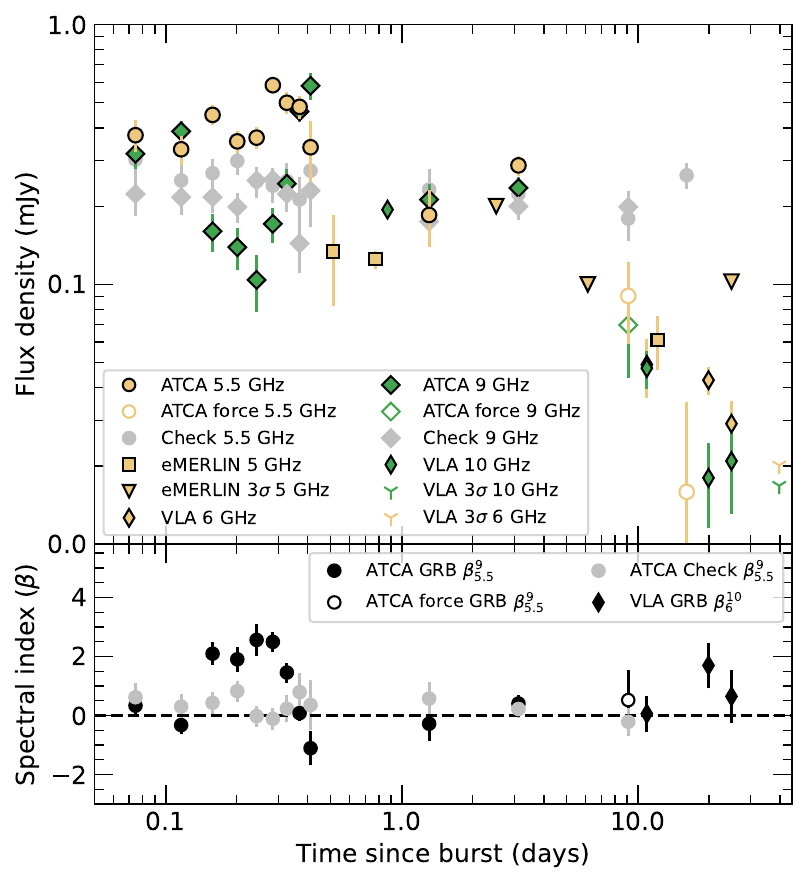}
    \caption{Top panel: The radio light curve of \thisgrb\ at 5--6\,GHz (yellow) and 9--10\,GHz (green) obtained with ATCA, \textit{e}-MERLIN and VLA \citep[the latter data points are taken from][]{schroeder25}. A nearby check source detected at both frequencies in the ATCA images is also plotted in grey to demonstrate the observed variability of \thisgrb\ is real. The solid data points show ATCA detections ($>3\sigma$). In contrast, the open data points are force-fitted flux densities at the known position of the source in the ATCA images when there was no significant detection. \textit{e}-MERLIN and VLA non-detections are plotted as $3\sigma$ upper limits. Bottom panel: Spectral index evolution (where the flux density is $S_{\nu} \propto \nu^{-\beta}$) between the ATCA 5.5 and 9\,GHz detections of the GRB (black) and check source (grey). An open data point is plotted when at least one of the flux densities used to calculate the spectral index was force-fitted in the ATCA image. The spectral evolution from the VLA 6 and 10\,GHz detections are also plotted as black diamonds.}
    \label{fig:lc_spec}
\end{figure*}

The full light curve of \thisgrb\ at 5.5 and 9 GHz can be found in the top panel of Figure~\ref{fig:lc_spec}, with the first observation split into 1-hour timebins. All plotted flux densities for the GRB source were calculated using the {\sc Miriad} task {\sc uvfit}, which fits for a point source near a defined position in the $uv$-plane. Before running the {\sc uvfit} algorithm, all other sources in the field were subtracted from the visibilities using the clean model. The same technique was applied to measure the flux densities of a nearby check source, which was used to characterise the variability of the afterglow. 
The check source was detected at both 5.5 and 9\,GHz and is located at RA (J2000): $22^{\rm h} 09^{\rm m} 38\fs23 \pm 2\farcs1$, Dec (J2000): $+13\degr 30' 36\farcs1 \pm 2\farcs1$, 83\,arcsec from \thisgrb.
In the case where {\sc uvfit} did not converge, the position of the GRB or check source was fixed, forcing the algorithm to fit a point source at the known source location. These force-fitted flux densities are plotted as open data points rather than filled data points in Figure~\ref{fig:lc_spec}. 
The reported flux density errors are the output {\sc uvfit} statistical error and a 5\% absolute flux density calibration error added in quadrature. The measured and force-fitted flux density measurements and $3\sigma$ upper limits are listed in Table~\ref{tab:radio_flux}.

\subsubsection{MeerKAT}\label{sec:meerkat}

The MeerKAT radio telescope is a 64-dish interferometer in the Karoo Desert, South Africa. Each dish is 13.5\,m in diameter and with a maximum baseline of 8\,km, MeerKAT can achieve an angular resolution of $\sim4-5$\,arcsec with the $L$-band ($856–1711$\,MHz) receiver. Four observations of \thisgrb\ were obtained through a Director's Discretionary Time proposal (PI: Rhodes, DDT-20231124-SA-01) at a central frequency of 1.3\,GHz with a bandwidth of 0.86\,GHz. Each dataset was processed using \textsc{oxkat} \citep{2020ascl.soft09003H}. \textsc{oxkat} is a series of semi-automated python scripts that wrap around \textsc{casa}, \textsc{tricolour}, \textsc{wsclean}, and \textsc{cubical} commands to reduce the data \citep{2007ASPC..376..127M,2014MNRAS.444..606O,2018MNRAS.478.2399K,2022ASPC..532..541H}. A given dataset is averaged down to 1024\,channels and 8\, second integrations. The calibrator scans are flagged, then bandpass, flux density and complex gain calibration are performed using PKS B1934-638 (the flux density, bandpass and delay calibrator) and J2232+1143 (gain calibrator). The calibration solutions were then applied to the target field, which was subsequently flagged and imaged. To further reduce the image rms noise, we performed a single round of phase-only self-calibration. 

We detected a counterpart to \thisgrb\ in two of the four observations. We measured the flux densities using \textit{imfit} within \textsc{casa} \citep{2007ASPC..376..127M}. The uncertainties associated with the flux density measurements are a combination of the statistical uncertainty and a 10\% absolute flux density calibration error added in quadrature. We provide the observation times, measured and force-fitted flux densities and $3\sigma$ upper limits in Table~\ref{tab:radio_flux}. 
These MeerKAT observations were also processed independently of  \citet{schroeder25}.

\subsubsection{AMI--LA}\label{sec:ami}

The Arcminute Microkelvin Imager -- Large Array (AMI--LA) is a facility of eight 12.8\,m dishes based at the Mullard Radio Astronomy Observatory, Cambridge, UK. Observations are made at a central frequency of 15.5\,GHz with a bandwidth of 5\,GHz. The position of \thisgrb\ was observed twice with the AMI--LA at 0.6 and 9.6\,days post-burst. The data were 
reduced using \textsc{reduce\_dc}, a custom software package \citep{2013MNRAS.429.3330P}, which performs flux density and complex gain calibration. 
The flagging, interactive cleaning, and imaging were performed in \textsc{casa} \citep{2007ASPC..376..127M}. In neither observation do we find a counterpart to \thisgrb\ and report our 3$\sigma$ upper limits in Table~\ref{tab:radio_flux}. We used the \textsc{casa} viewer tool \citep{2007ASPC..376..127M} to measure the rms noise in each image from which the 3$\sigma$ upper limits were calculated.

\subsubsection{\textit{e}-MERLIN}\label{sec:emerlin}

The \textit{enhanced} Multi-Element Radio-Linked Interferometer Network (\textit{e}-MERLIN) is a UK-based very long baseline interferometry facility. With a maximum baseline of 217\,km, \textit{e}-MERLIN can obtain an angular resolution of 0.05 arcsec at 5\,GHz. Observations of \thisgrb\ were obtained through a successful open-time Target of Opportunity proposal (PI: Rhodes, CY16204). Of the five observations, the third, fourth, and fifth epochs used the Lovell because of its new frequency flexibility capabilities.
\textit{e}-MERLIN observations were reduced with a custom \textsc{casa}-based pipeline \citep{2021ascl.soft09006M}. The pipeline performs flagging, flux density scaling (using J1331+3030) followed by two iterations of bandpass and complex gain calibration (using J1407+2827 and J2218+1520, respectively). Images are made using iterative cleaning and deconvolution. 

We detected the radio counterpart to \thisgrb\ in two of the five observations (epochs 1 and 4).
For our first detection, which was in the first epoch, the high signal-to-noise ratio enabled us to split the observation in time to search for intra-observation variability like that observed with ATCA. 
The flux densities were measured using \textit{imfit} within \textsc{casa} \citep{2021ascl.soft09006M}. Our \textit{e}-MERLIN detections and upper limits are plotted in Figure~\ref{fig:lc_spec} and are listed in Table~\ref{tab:radio_flux}.

\subsection{Optical Observations}\label{sec:optical}

\subsubsection{SLT optical afterglow discovery}\label{sec:slt}

\begin{figure}
    \centering
    \includegraphics[width=\columnwidth]{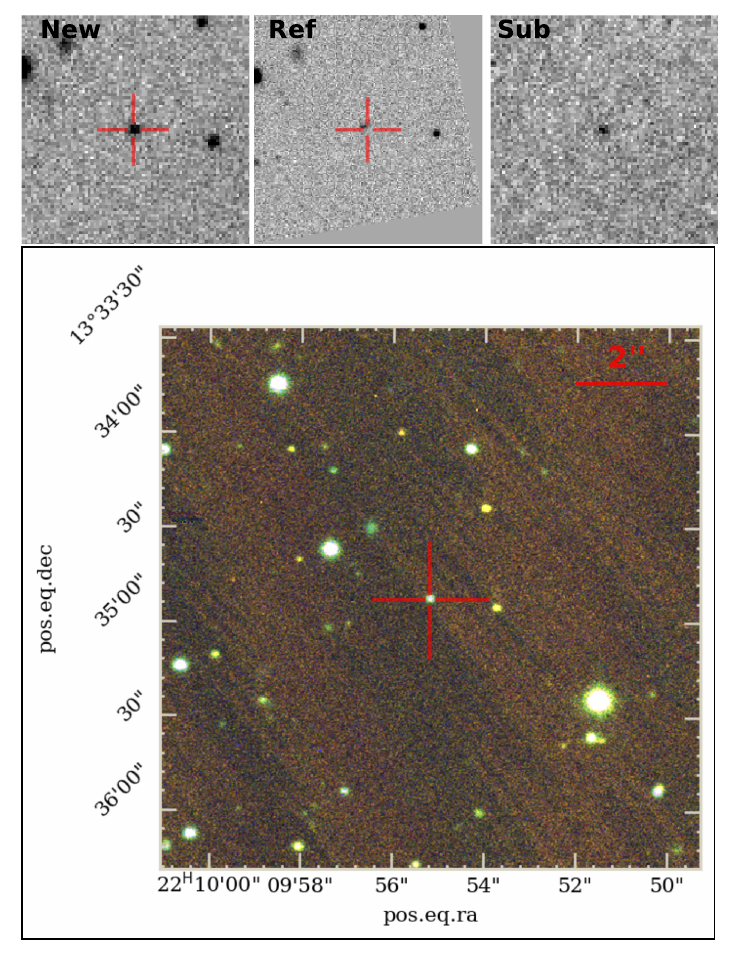}
    \caption{Discovery of the \thisgrb\ optical afterglow. Upper: 40cm-SLT $r$-band discovery image taken 7.39 hours after the Swift trigger (left), subtracted from the SDSS template (middle) and afterglow identification (right). Bottom: LOT gri-colour combined image of the afterglow and field.}
    \label{fig:discovery}
\end{figure}

We used the 40cm SLT telescope at Lulin Observatory, Taiwan, to obtain $r$-band images for the \thisgrb\ field as part of the Kinder collaboration \citep{chen25}. 
The afterglow discovery reported by \citet{yang23GCN} coincides with a faint (Petrosian magnitude of $r=21.32\pm0.22$\,mag) galaxy, SDSS J220933.34+133119.5.
For the discovery image of the optical afterglow see Figure~\ref{fig:discovery}.
We continuously observed the afterglow with SLT and found a plateau within the $r$-band light curve \citep{2023GCN.35105....1C}. 
We employed the Kinder pipeline \citep{kinderpip} to conduct point spread function (PSF) photometry for the afterglow after template subtraction using the SDSS images. The magnitudes and a $3\sigma$ detection limit in the AB system are reported in Table~\ref{tab:optical}.
We registered the astronomical transient name AT~2023yba \citep{chen23TNS_at23yba} when the object was not yet confirmed as the afterglow in the early stages.

\subsubsection{CAHA}

The Calar Alto Faint Object Spectrograph (CAFOS) mounted on the 2.2~m telescope at the Spanish Astronomical Center in Andalusia (Almería, Spain) was used to observe GRB\,231117A starting at 17:47:16 UT on 2023 November 17. The observation was performed in $i$ band and consisted of 14 exposures with a total exposure time of 3390 s. Images were reduced in a standard fashion, including bias, flat field and astrometric corrections using IRAF \citep{1986SPIE..627..733T} and astrometry.net \citep{2010AJ....139.1782L}. Photometry of this observation is provided in Table~\ref{tab:optical}.

\subsubsection{GOTO}\label{sec:goto}

The Gravitational-wave Optical Transient Observer \citep[GOTO;][]{Steeghs22, Dyer24} responded autonomously to the \swift-BAT trigger and began pointed observations from its southern node in Siding Spring Observatory, Australia, at 10:00:02 UT on 2023 November 17, $7.0$ hours after trigger. Observations were performed in the GOTO $L$ filter ($400-700$\,nm), roughly equivalent to the SDSS $g$ + $r$ filters. A total of six epochs were obtained over the first three nights from GOTO-North (La Palma, Spain) and GOTO-South (Siding Spring Observatory, Australia). Images were processed using the GOTO pipeline, including image alignment and template subtraction using recent pointings of the same field. Magnitudes are calibrated against ATLAS Refcat2 \citep{Tonry18} and are presented in Table~\ref{tab:optical}.

\subsubsection{GTC}
The 10.4~m Gran Telescopio Canarias (GTC) observed the field of GRB\,231117A 
using EMIR \citep{2022A&A...667A.107G} in the infrared and OSIRIS+ \citep{2000SPIE.4008..623C} in the optical. The EMIR observation started on 2023 November 19 at 20:01:31.63 UT with a total exposure time of 1617 s in the $K_S$-band. 
OSIRIS+ performed a $z$-band observation of the field on 2023 November 25 at 20:32:39 UT with a total exposure time of 600~s (note it was affected by the full Moon). 
All data were reduced using standard techniques, including bias subtraction, flat-field correction, fine alignment, stacking and astrometric calibration.
These observations are listed in Table~\ref{tab:optical}.

\subsubsection{Liverpool Telescope}\label{sec:LT}

We triggered target-of-opportunity observations with the 2-m robotic Liverpool Telescope \citep[LT;][]{Steele04} under programme number PL23B19 (PI: Gompertz). Observations were performed using the $r$ and $i$ filters \citep{Fukugita96} over three nights, starting at 20:05:45 UT on 2023 November 17, $0.69$ days following \swift-BAT trigger. Photometry was performed using Photometry Sans Frustration\footnote{\href{https://github.com/mnicholl/photometry-sans-frustration}{https://github.com/mnicholl/photometry-sans-frustration}} \citep{Nicholl23}, including image alignment and stacking within individual nights, and template subtraction using Panoramic Survey Telescope and Rapid Response System (Pan-STARRS) reference images \citep{Flewelling2020} to remove host galaxy light. Magnitudes are reported in Table~\ref{tab:optical}. 

\subsubsection{LOT}\label{sec:lot}

We also conducted a follow-up campaign using the Lulin One-meter Telescope (LOT) with multi-band imaging in $g, r, i$, and $z$ during the first epoch \citep{yang23GCN}, as part of the Kinder collaboration \citep{chen25}. The strategic advantage of LOT being located at the same site as SLT, coupled with its relatively larger diameter, contributed to the multi-colour information and late-time deep imaging of GRB~231117A. The data reduction process and photometric measurements followed the same procedures as the SLT data, and the template-subtracted magnitudes are detailed in Table~\ref{tab:optical}. 

\subsubsection{NTT}
We observed the field of GRB\,231117A with the European Southern Observatory (ESO) Faint Object Spectrograph and Camera \citep[v.2, EFOSC2,][]{efosc_buzzoni+84} mounted on the ESO New Technology Telescope (NTT) at La Silla Observatory, Chile, under the ePESSTO+ programme. Observations were obtained in the $R$ filter starting from 2023 November 18 at 00:22:34, i.e. 0.884 days after the trigger. Further observations were performed at later epochs at 6.92 and 8.90 days.
We performed image subtraction using templates from Pan-STARRS images \citep{Flewelling2020} with the \texttt{HOTPANTS} \citep[High Order Transform of Psf ANd Template Subtraction code,][]{HOTPANTS} package to remove the host galaxy light. Aperture photometry was performed using \texttt{SExtractor} \citep{Sextractor}, and the results are reported in Table~\ref{tab:optical}.

\subsubsection{Pan-STARRS}\label{sec:panstarrs}

Pan-STARRS consists of two 1.8-metre telescope units located at the summit of Haleakala on the Hawaiian island of Maui \citep{Chambers2016arxiv}. We triggered Pan-STARRS1 (PS1) to observe in $grizy_{\textsc{ps}}$, and a total of 19~epochs were taken across the 18th and 19th of November, with observations starting at approximately 06:08 and 06:35~UT, respectively. Photometry was carried out on the difference images with the Pan-STARRS1 $3 \pi$ sky survey data used as references, and to set photometric zero points on the target images. We used $200 - 300$\,s exposures and stacked the PSF flux measured on the difference images for each filter, using daily weighted averages to enhance the signal-to-noise ratio (SNR). The magnitudes derived from these flux stacks are listed in Table~\ref{tab:optical}.

\subsubsection{VLT}\label{sec:vlt}

We observed the afterglow using FORS2 and HAWK-I mounted the Very Large Telescope (VLT), under the Stargate program.
FORS2 observations started on 2023-November 19, 1.9 days after the burst trigger, and were obtained in the $g_{\mathrm{HIGH}}$, $R_{\mathrm{SPECIAL}}$, $I_{\mathrm{BESS}}$, $Z_{\mathrm{GUNN}}$ broadband filters. We secured further imaging 3.9 and 8.9 days after the trigger, respectively using the $R_{\mathrm{SPECIAL}}$- and $Z_{\mathrm{GUNN}}$- bands.
HAWK-I observations also started on 2023-November 19, 1.9 days after the burst trigger, and were obtained in the $J$, $H$, and $K$ broadband filters. We secured further imaging 4.9 and 6.9 days after the trigger, as well as approximately 1 month later on 2024 December 12 to obtain a template image.

FORS2 and HAWK-I data were reduced using the ESO Reflex environment \citep{esoreflex2013a}. To remove the fringing in the FORS images, especially strong in the first epoch (except in $g$-band) due to the use of the blue CCD, we used fringing patterns obtained from the same dithered images of the field. We applied aperture photometry with a curve of growth to measure the flux of the combined host-galaxy and afterglow. 
To remove the contribution of the host, we applied image subtraction using a routine based on \texttt{HOTPANTS}
, and archival Pan-STARRS images as reference. Afterwards, PSF photometry was used to measure the magnitudes of the GRB afterglow. 
The FORS2 and HAWK-I magnitudes are listed in Table~\ref{tab:optical}.

\subsubsection{Host observations}\label{sec:obs_host}

To explore the host galaxy, we first retrieved science-ready coadded images from the DESI Legacy Imaging Surveys \citep[Legacy Surveys, LS;][]{Dey2018a} data release (DR) 10 and the Panoramic Survey Telescope and Rapid Response System (PS) DR1 \citep{Chambers2016arxiv}. We augmented these data with $K$-band imaging from our HAWK-I campaign (Section~\ref{sec:vlt}) and archival $u$-band images obtained with MegaCAM at the 3.58\,m Canada-France-Hawaii Telescope (CFHT). We measured the brightness of the host using  {\sc LAMBDAR} \citep[Lambda Adaptive Multi-Band Deblending Algorithm in R;][]{Wright2016a}, the methods described in \citet{Schulze2021a} and the $K$-band photometry with an aperture photometry tool from \citet[][and references therein]{Schulze2018a}. The instrumental magnitudes are calibrated against stars from 2MASS, Legacy Survey and Pan-STARRS. Table \ref{tab:phot:host} presents the measurements of the host galaxy in the different bands.

\subsection{\swift\ Observations}\label{sec:swift}

\subsubsection{UVOT}\label{sec:uvot}

The {\it Swift} Ultraviolet and Optical Telescope \citep[UVOT;][]{roming04} began settled observations of the field of GRB 231117A
106 s after the \swift-BAT trigger \citep{kuin23}. Due to issues with the gyroscopes onboard {\it Swift}
at the time of observations \citep{cenko23}, the UVOT images were slightly blurred due to the degraded attitude 
control of the spacecraft; this only affected the UVOT image quality as UVOT has finer angular resolution than XRT and BAT.
Due to the blurring, source counts were extracted using a source region of 7 arcsec radius. To be consistent
with the UVOT calibration, these count rates were then corrected to a 5 arcsec radius using the curve
of growth contained in the calibration files. Background counts were extracted using a circular region with a 20 arcsec
radius located in a source-free region. The count rates were obtained from the image lists using the {\it Swift} tool {\sc uvotsource}. 
At late times, the light curves were contaminated by the underlying host galaxy. 
To estimate the level of contamination, for the $white$, $u$, $uvw1$ and $uvw2$ filters we combined the late-time
exposures \citep{kuin23b} beyond 10$^7$s until the end of observations. We extracted the count rate in the late combined exposures using
the same 7 arcsec radius aperture, which was then corrected to a 5 arcsec radius using the curve of growth contained in the calibration files. 
These were subtracted from the source count rates derived with the same size aperture to obtain the afterglow count rates. 
The afterglow count rates were converted to magnitudes using the UVOT photometric zero points \citep{poole08,breeveld11}.

\subsubsection{XRT}\label{sec:xrt}

{\it Swift}-XRT observations of \thisgrb\ began 93\,s following the \swift\ detection and continued until 4\,days post-burst \citep{laha23gcn,beardmore23,melandri23}. The XRT observations were downloaded from the UK \emph{Swift} Science Data Centre (UKSSDC) \citep{Evans07,Evans09}. All data that were flagged as `bad data points' were removed. Data were rebinned to limit orbit gaps to 10\,ks. To find the counts-to-unabsorbed-flux conversion factors of the rebinned data, we fit spectra in four phenomenologically-identified regions: $93$ -- $300$s; $300$ -- $2000$s; $2000$ -- $10^4$s; and $10^4$ -- $2 \times 10^6$s. The derived conversion factors were $(4.19 \pm 0.21) \times 10^{-11}$\,erg\,cm$^{-2}$\,ct$^{-1}$, $(3.95 \pm 0.14) \times 10^{-11}$\,erg\,cm$^{-2}$\,ct$^{-1}$, $(4.47 \pm 0.26) \times 10^{-11}$\,erg\,cm$^{-2}$\,ct$^{-1}$, and $(3.39 \pm 0.21) \times 10^{-11}$\,erg\,cm$^{-2}$\,ct$^{-1}$, respectively. Pile-up correction was also applied.

\section{Wavelength Dependent Analysis} \label{sec:wave}

\subsection{Radio Variability}\label{sec:rad_var}

The radio light curves and spectral evolution of \thisgrb\ as observed by ATCA (5.5 and 9\,GHz), \textit{e}-MERLIN (5\,GHz) and the VLA \citep[6 and 10\,GHz from][]{schroeder25} are plotted in Figure~\ref{fig:lc_spec}.  
The light curves in the top panel are grouped into two frequency bands: 5--6\,GHz and 9--10\,GHz. 
The first ATCA observation was split into nine 1-hour time bins, and the first \textit{e}-MERLIN observation split into two 6-hour time bins before plotting.
The bottom panel of Figure~\ref{fig:lc_spec} shows the evolution of the spectral index between the simultaneous ATCA 5.5/9\,GHz and VLA 6/10\,GHz observations. 
We also plot the light curves and the spectral evolution of the check source as gray data points in Figure~\ref{fig:lc_spec} to test the observed variability of the radio afterglow. 
We performed a $\chi^{2}$ test on the check source following the method outlined by \citet{bell14}, confirming it to be non-variable across all ATCA observations with a probability of $P>0.99$ at both frequencies. 

We searched for short-term intra-observational variability of the radio afterglow within the first and third ATCA observations ($<10$\,hr and 3.1\,days post-burst) on 1-hour timescales (the second ATCA observation at 1.3\,days suffered from poor observing conditions, which prevented a similar analysis). 
We calculated the $\chi^{2}$ probability that the GRB remained constant throughout the observation, defining a source as variable if this probability was $P<1 \times 10^{-3}$ \citep{bell14}. 
In the first ATCA observation, the probability that the GRB was a steady source was $P=1.7 \times 10^{-4}$ and $P=1.4 \times 10^{-24}$ at 5.5 and 9\,GHz, respectively, confirming it as variable. In comparison, the check source was $P=0.9$ and $P=0.4$ within that same observation. In the third ATCA observation, the probability that the GRB and check source were steady sources were all $P>0.4$ at 5.5 and 9\,GHz. We therefore conclude that the GRB was variable on 1\,hr timescales during the first ATCA observation but not during the third ATCA observation 3.1\,days post-burst.

We also search for intra-observational variability in the first 12-hour \textit{e}-MERLIN observation, which was split into two 6-hour observations centered at 0.49 and 0.76\,days, with the resulting flux densities being consistent between these times.  
We did a similar analysis of the first MeerKAT observation at 9.5\,days, splitting the observation into four $\sim30$\,min time bins. No variability was detected with a probability of $P=0.6$ of being steady during the 2-hour observation. 

While the first ATCA observation shows short term variability, the overall light curve trend at both 5--6 and 9--10\,GHz shows the afterglow evolution to be reasonably flat until $\sim3$\,days post-burst, at which time it began to decay. Such behavior is indicative of a plateau. 

\subsection{Scintillation and source size estimate}\label{sec:scint}

The observed variability in the first ATCA observation is likely attributed to interstellar scintillation (ISS) as has been observed from other GRB radio afterglows \citep[e.g.][]{frail97,frail00,waxman98,chandra08,vanderhorst14,gompertz17,alexander19,rhodes22,anderson23} but never at such early times. 
The observed modulation index is calculated according to $m=\sigma_s/\bar{S}$, where $\sigma_s$ is the standard deviation in the flux densities of the GRB and $\bar{S}$ is the mean flux density.
We estimated the variability timescales of \thisgrb\ at both frequencies during the first ATCA observation based on the perceived modulation of the light curves. The 5.5\,GHz flux density measurements appear to modulate between adjacent data points on hourly timescales. Meanwhile, the 9\,GHz measurements appear to modulate once over the full 9-hour observation.  

The observed modulation indices and timescales are listed in Table~\ref{tab:scint_pred} and are directly compared to the theoretical values calculated following the relations in \citet{goodman97} and \citet{walker98}.
We used the Python wrapper {\sc pyne2001}\footnote{https://github.com/v-morello/pyne2001}, which uses the NE2001 model of the Galactic distribution of free electrons \citep{cordes02} to derive a transition frequency of $\nu_{0}=10.1$\,GHz and a scattering measure of $SM=2.1 \times 10^{-4}$\,kpc\,m$^{-20/3}$.
This places both the 5.5 and 9\,GHz observing frequencies in the strong scintillation regime. 
Using these parameters, we list the size of the scattering disk $\theta_{\mathrm{scatt}}$, predicted modulation index ($m_{\mathrm{p}}$) and scattering timescale ($t_{\mathrm{scint,p}}$) for both the strong refractive and diffractive regimes in Table~\ref{tab:scint_pred}, assuming \thisgrb\ is a point source at 5.5 and 9.0\,GHz $\lesssim10$\,hr post-burst. 

While lower than the theoretical values, our measured modulation indices for \thisgrb\ do support scintillation being detected in the first ATCA observation at both frequencies $\lesssim10$\,hr post-burst.
Given 9\,GHz is closer to the transition frequency, we expect it to scintillate more than at 5.5\,GHz. 
Meanwhile, our estimated timescales are different to the predicted values.
However, we note that the theoretical estimates assume a certain scattering screen distance so different distances can lead to different timescales and modulation indices.  
As the radio afterglow expands, we expect the scintillation effect to decrease, but more slowly at lower frequencies, where the scattering scales are largest. 
While scintillation affected the first ATCA observation $\lesssim10$\,hr post-burst, the afterglow had expanded above the scattering scale at 5.5\,GHz by $\sim3.1$\,days and 1.3\,GHz by $\sim9.5$\,days. 

Assuming the first ATCA observation is experiencing scintillation, we can use the theoretical scattering scale $\theta_{\mathrm{scatt}}$ predictions in Table~\ref{tab:scint_pred} to place an upper limit on the size of the blast wave at the angular size distance $D_{\mathrm{A}}$ derived from the redshift $z=0.257$ \citep{gonzalex-ba23GCN}. 
We cannot distinguish between strong refractive and diffractive regimes using these data so we assume the larger scattering scale of strong refractive scintillation for this calculation.
We predict a maximum radial size of $1\times10^{16}$\,cm (expected uncertainties likely $\sim50$\%) at 9\,GHz, consistent with the range derived from scintillation of other GRBs \citep[$\sim1\times 10^{16} \mathrm{~to~} \sim8 \times 10^{16}$\,cm;][]{frail97,frail00,waxman98,chandra08,vanderhorst14,alexander19,rhodes22,anderson23}.
This is the earliest (1.7\,hours post-burst) source size estimate derived from scintillation for any GRB to date.

\begin{table*}
    \centering
    \begin{tabular}{cccclcccc}
         \hline
         $\nu$ &  \multicolumn{2}{c}{Measured} &  \multicolumn{5}{c}{NE2001 model Predictions}  \\
         \cline{2-3} \cline{5-9} 
          & $m$ & $t_{\mathrm{scint}}$ & & Regime & $\theta_{\mathrm{scatt}}$ & $m_{\mathrm{p}}$ & $t_{\mathrm{scint,p}}$ & Size  \\
         (GHz) &  & (hr) & & & ($\mu$as) & &(hr) & ($10^{16}$\,cm)  \\
         \hline
         \hline
         5.5  & 0.2  &  1 & & R & 3.1 & 0.7 & 7.6 & $\lesssim4$  \\
              &      &   & & D & 0.4 & 1.0 & 1.0 & \\
        \hline 
         9.0  & 0.5 & 9  & & R  & 1.0 & 0.9 & 2.6 & $\lesssim1$ \\
              &      &   & & D & 0.7 & 1.0 & 1.7 & \\
    \end{tabular}
    \caption{The modulation index ($m$) and scintillation timescale ($t_{\mathrm{scint}}$) measured from the first ATCA observation where the 5.5 and 9\,GHz data were split into 1-hour time bins. These are compared to scintillation predictions from NE2001 according to strong refractive (R) and diffractive (D) scintillation regime relations \citep[summarised in table 1 of][]{granot14}.
    }
    \label{tab:scint_pred}
\end{table*}

\subsection{Optical extinction}\label{sec:extinct} 

To estimate extinction in the host galaxy, we extracted optical and near infrared (NIR) SEDs using the Pan-STARRS $grizy$ data centred at the two epochs of $\sim$1.1 and $\sim$2.1 days after the GRB trigger (see Table~\ref{tab:optical}). 
Fits were performed by adapting the photometric redshift estimation software package \texttt{phozzy} \citep{Fausey2023}. The code assumes a single power law through the optical-NIR regime, accounts for intergalactic attenuation by applying a k-correction to each photometric point following the prescription of \citet{Meiksin2006}, and for host galaxy extinction using models from \citet{Pei1992}. We used the Small Magellanic Cloud (SMC) extinction model for our fit because it best represents observations of GRB host galaxies \citep{Schady2012}. Pan-STARRS photometric band data were corrected for Milky Way extinction using the NASA/IPAC Extragalactic Database (NED) Extinction Calculator\footnote{The NASA/IPAC Extragalactic Database (NED) is funded by the National Aeronautics and Space Administration and operated by the California Institute of Technology.} and \citet{Schlafly11}, and then fit for the normalization at $\sim 0.96$ $\mu$m ($A$), the spectral index ($\beta$), and the host galaxy extinction ($E_{\rm B-V}$), using a Markov Chain Monte Carlo (MCMC) implementation of the \texttt{emcee} software package \citep{ForemanMackey2013}. The redshift was fixed to $z = 0.257$.

We used a broad Gaussian spectral index prior centred on $\beta_{\mathrm{O}} = 0.81$ with a standard deviation of $\sigma = 0.20$, based on the time-averaged photon index reported in the \emph{Swift}-XRT Spectrum Repository \citep{Evans07,Evans09}, and an extinction prior based on the GRB host galaxy extinction distribution from \citet{Covino2013}. At $\sim 1.1$ days we find a normalization of $A = 30.71^{+4.31}_{-2.95}$ $\mu$Jy, a spectral index $\beta_{\mathrm{O}, ~1.1~\mathrm{days}} = 0.68^{+0.17}_{-0.16}$ and a 1$\sigma$ (3$\sigma$) host galaxy extinction upper limit of $E_{\rm B-V} <0.06 (<0.25)$. At $\sim 2.1$ days we find a normalization of $A = 11.02^{+2.46}_{-1.53}$ $\mu$Jy, a spectral index $\beta_{\mathrm{O}, ~2.1~\mathrm{days}} = 0.79^{+0.18}_{-0.19}$ and a 1$\sigma$ (3$\sigma$) host galaxy extinction upper limit of $E_{\rm B-V} < 0.12(<0.34)$. We therefore conclude that the contribution of the GRB 231117A host galaxy to optical extinction is negligible.

\section{GRB Afterglow modeling}\label{sec:afterglow_model}

In the following section, we describe the modelling of the GRB afterglow, with the full multi-wavelength light curve of GRB 231117A shown in Figure~\ref{fig:powerlaws}. 
At early times, $\lesssim 0.02$\,days post burst,
\swift-XRT and -UVOT observations show a declining profile. 
This early decline is followed by an apparent plateau at optical and radio frequencies lasting up to $\sim1-3$\,days post-burst.
During this time, an X-ray flare is seen at $\sim 0.07$\,days and the (sparse) X-ray data are consistent with a shallower light curve decline than the one established at very early times. 
From $\sim 1$\,day, the multi-wavelength afterglow starts to decline again.
We first compare the temporal and spectral properties to standard closure relations to explore initial constraints on the afterglow. This is followed by details of the afterglow modelling.

\subsection{Afterglow behaviour}\label{sec:afterglow_model_1}

We compare the afterglow observations with the standard closure relations to determine consistency between the bands and the various temporal indices \citep{sari98, granot02}.
The spectral index between X-ray and optical data at $\sim 5$\,days is $\beta_{\mathrm{X/O}, ~5~{\rm days}}\sim0.8$. This is consistent with $\beta_{\mathrm{X},~<5~{\rm hours}}=0.68^{+0.34}_{-0.25}$ derived from XRT observations $<5$\,hr post-burst 
\citep{melandri23} and the $\beta_{\mathrm{O},~2.1~{\rm days}}=0.79^{+0.18}_{-0.19}$ at $2.1$\,days post-burst derived in Section~\ref{sec:extinct}, placing the X-ray and optical bands in the same regime.
The declining temporal lightcurve at this epoch indicates that the X-ray and optical frequencies are above the characteristic synchrotron peak frequency ($\nu > \nu_m)$, and in the same spectral regime, either $\nu<\nu_c$, where $\beta = (p-1)/2$ or $\nu > \nu_c$, where $\beta = p/2$.
Using our derived value for $\beta_{\mathrm{X/O}}$, these two possible spectral regimes allow for two possible values for $p$, the electron distribution index (such that $N_e \propto \gamma_e^{-p}$); $p = p_1 \sim 2.7$, if $\nu<\nu_c$; or $p = p_2 \sim 1.7$, if $\nu>\nu_c$.

The closure relations for these values of $p_{1,2}$ and their spectral regimes, assuming a uniform medium and slow cooling,\footnote{This is expected for SGRBs. Similarly, a wind medium can be ruled out with no viable combination of $\alpha$ and $\beta$, either fast or slow cooling, nor pre- or post-jet break.}
are $\alpha_1 = 3(p_1-1)/4$ for $\nu<\nu_c$ and $p>2$, and $\alpha_2 = (3p_2 + 10)/16$ for $\nu>\nu_c$ and $p<2$ \citep[see Tables 13 \& 15 respectively in][]{gao13}.
Using the values of $p_{1,2}$ derived from the $\beta_{\mathrm{X/O}}$ above, these give $\alpha_1 \sim 1.3$ and $\alpha_2 \sim 0.9$.
A linear regression power law fit to the optical data at $2-5$\,days, results in $\alpha_{\mathrm{O},2-5~{\rm days}} = 1.66 \pm 0.10$, which is not immediately consistent with either $\alpha_1$ or $\alpha_2$.
However, if we invoke a geometric jet break (edge effect only) at the start of this epoch at $\gtrsim 2$\,days, the case for $p_2\sim 1.7$ gives $\alpha_2 + 3/4 = 1.7$, which is equivalent to 
$\alpha_{\mathrm{O},2-5~{\rm days}} = 1.66 \pm 0.10$ \citep[note the addition of $3/4$ is a result of the edge effect on the post break temporal evolution; Table 18 in][]{gao13}.
For the two possible values for the electron distribution $p_1$ and $p_2$, we therefore favor $p = p_2 = 1.7$ and use this as our fiducial value going forward.

\begin{figure*}
    \centering\includegraphics[width=\textwidth]{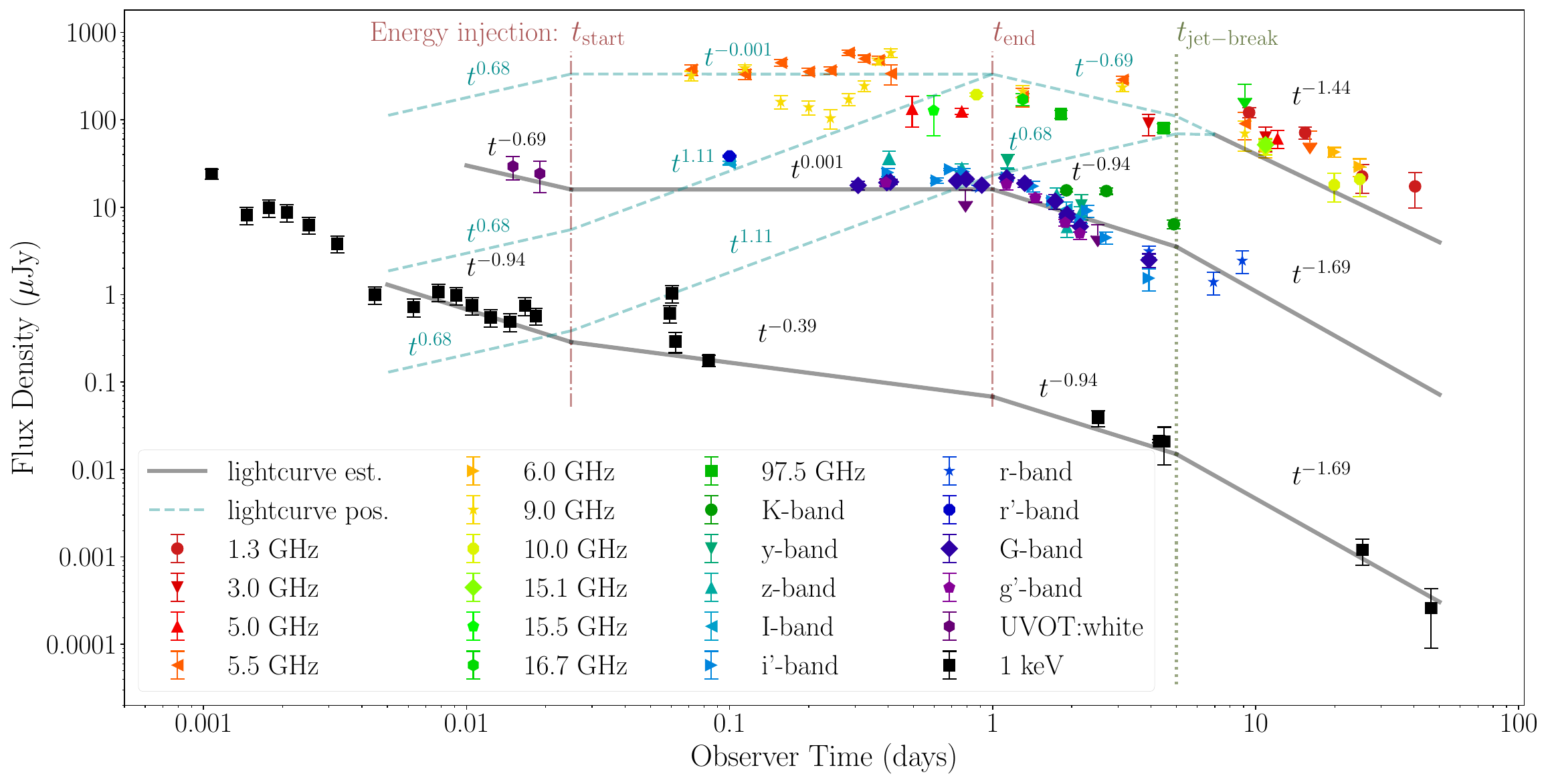}
    \caption{Estimating afterglow emission characteristics using closure relations vs the observed absorption/extinction corrected data. Our best reproduction requires a forward shock afterglow with $p<2$, and a period of significant energy injection, between $\sim0.02 - 1.0$\,days post-burst. Temporal indices for the various components are listed, as are the key timescales -- the $t^{-\alpha}$ labels correspond to the line immediately below the label. The solid grey lines are our lightcurve estimates (labelled ``lightcurve est."), and the dashed blue lines give possible radio lightcurve behaviours for before the jet break (labelled ``lightcurve pos.") dependent on the spectral regime at radio wavelengths, before or after the passage of the synchrotron self-absorption frequency.}
    \label{fig:powerlaws}
\end{figure*}

The observed temporal power law decline at $\gtrsim 1$\,day, but before a jet break at $2 \lesssim {\rm jet~ break} \lesssim 5$\,days, is consistent at X-ray and optical wavelengths with $\alpha_{\mathrm{X/O},~>1~{\rm day}}\sim0.9 \sim \alpha_2$ and would require the radio emission to be either rising with an index $\alpha_{2,~{\rm R}} = -(p_2+2)/[8(p_2-1)] \sim -0.7$ (if $\nu<\nu_m$) or following a shallow decline with $\alpha_{2,~{\rm R}} = 3(p_2+2)/16 \sim 0.7$ (if $\nu>\nu_m$). 
When also considering the temporal decline observed at radio frequencies from $\sim 1-3$\,days $\alpha_{\mathrm{R},~1-3~{\rm days}}\sim 0.7$, which steepens at $\gtrsim5$\,days to $\alpha_{{\rm R},~>5~{\rm days}}\sim 1.4$, the multi-wavelength data are also indicative of a geometric jet break with $\alpha_{2,{\rm ~R}} + 3/4 \sim 1.4 \equiv \alpha_{{\rm R},~>5~{\rm days}}${\rm ; where the relations are consistent with our fiducial} value for $p = p_2$, and the index $\alpha_{2,~{\rm{X/O}}}\sim 1.7$ at optical and X-ray frequencies (see Figure\,\ref{fig:powerlaws}, where self-consistent closure relation temporal indices are shown for illustrative purposes).

The pre-break ($\lesssim 2$ to $5$\,days) optical and X-ray declines are consistent after 1 day with an $\alpha_{\mathrm{X/O},~>1~{\rm day}} \sim 0.9$. 
This is also consistent with the observed X-ray decline at $\lesssim 0.02$\,days, $\alpha_{\rm X,~<0.02~days}\sim 0.9$, and indicates that the X-rays are always in the same spectral regime above the cooling break, $\nu_X>\nu_c$. However, the very early optical data at $< 0.02$ days has a decline index of  $\alpha_{\rm O,~<0.02~days} \sim 0.65$ (again calculated using a linear regression power law fit); this shallower decline points to a different spectral regime for the optical emission at this time.  
If at these early times the cooling break is between the optical and the X-rays, $\nu_{\rm O} < \nu_c < \nu_{\rm X}$, then the change in the spectral index at $\nu_c$ is $\Delta\beta = 0.5$, resulting in an early optical decline such that $\alpha_{\rm X} - 1/4 = \alpha_{\rm O} = 0.65 \equiv \alpha_{\rm O,~<0.02~days}$, reproducing the observed decline.

The temporal decline in the X-rays at $<0.02$\,days and $>1$\,day (up until the jet break) are consistent. However, this $<0.02$\,day and $>1$\,day decline rate is not continuous between these times, where the optical and X-ray lightcurves show a significantly shallower decline.
To smoothly connect the early data at $t<0.02$\,days, with that at $t>1$\,day, requires a period of energy injection.
By solving the closure relations for the case of energy injection, a uniform medium, and $p<2$ \citep[see the right columns of Table 15][]{gao13} for the interval $0.02\lesssim t_{\rm inj} \lesssim 1$\,days, the required rate of energy injection is given by $E \propto \gamma^{1-s}$, where we fix $s = 2.84$ for our estimated lightcurve \citep{rees98}.
This is equivalent to a luminosity of $ L\propto t^{-q}$, with $q=0.44$ \citep{zhang06}\footnote{The injection closure relations in \citet{gao13} use $q$ to define the energy injection indices}.

The lightcurve estimates shown in Figure\,\ref{fig:powerlaws} provide a phenomenological explanation for the observed afterglow, where the uncertainty on the exact location of the various break frequencies (e.g., cooling, synchrotron, and self-absorption) leads to a broad range of potential behaviour, particularly at radio frequencies and at $\lesssim 1$\,day (see the blue dashed lines in Fig.\ref{fig:powerlaws}).
The closure relation study demonstrates that the data requires a period of energy injection into the afterglow between $\sim 0.02$ and $\sim 1$\,day post burst.

\subsection{Modeling}\label{sec:main_model}

Using the {\sc Redback} Bayesian inference package for electromagnetic transients \citep{sarin2024}, we fit an afterglow model to the data.
Motivated by the closure relation estimates, we used the \texttt{tophat-redback-refreshed} model \citep[see, ][]{lamb19, sarin2024}, within a uniform density environment and without lateral spread to ensure a geometric jet break.
We maintain broad prior ranges for the model parameters, as shown in Table \ref{tab:priors}, and utilise {\sc Nessai}, a nested sampling algorithm with artificial intelligence \citep[see, e.g.,][]{Williams:2021qyt, williams23nested}.
The \texttt{tophat-redback-refreshed} model includes a period of energy injection, with the additional free parameters: the Lorentz factor of the impulsive shell when energy injection begins ($\gamma_{\rm{collison}}$, a collision Lorentz factor); the factor by which the energy of the outflow increases at the end of the energy injection ($\zeta$); and the rate at which mass, or energy, is injected ($s$).
These parameters can be chosen so that no energy injection occurs (i.e., $\zeta = 1$, and $s = 1$).

\begin{table}[]
    \centering
    \begin{tabular}{c|c|c|c}
        Parameter & Prior range & Posterior & Unit \\
        \hline 
        $\log{E_0}$ & $48 - 54$ & $49.88^{+0.29}_{-0.18}$ & $\log$ erg \\
        $\theta_j$ & $0.02 - 0.45$ & $0.29^{+0.02}_{-0.02}$ & radian \\
        $\gamma_{\rm collision}$ & $3 - 30$ & $18.5^{+2.8}_{-1.9}$ & -- \\
        $\zeta$ & $1 - 100$ & $17.7^{+1.4}_{-1.3}$ & $E_{\rm final} / E_0 =\zeta$ \\
        $s$ & $0 - 10$ & $6.55^{+2.15}_{-1.74}$ & $E \propto \gamma^{1-s}$ \\
        $\log{n_0}$ & $-6 - 3$ & $0.63^{+0.32}_{-0.29}$ & $\log$ cm$^{-3}$ \\
        $p$ & $1.3 - 3.4$ & $1.66^{+0.01}_{-0.01}$ & $N_e \propto \gamma_e^{-p}$\\
        $\log{\varepsilon_e}$ & $-6 - -0.5$ & $-0.76^{+0.18}_{-0.29}$ & -- \\
        $\log{\varepsilon_B}$ & $-6 - -0.5$ & $-1.33^{+0.23}_{-0.31}$ & -- \\
        $\Gamma_0$ & $30 - 500$ & $405^{+67}_{-93}$ & -- \\
        $\Xi_N$ & $0.01 - 1.0$ & $0.07^{+0.03}_{-0.03}$ & --
    \end{tabular}
        \caption{The model parameters are defined in the following.\\ $E_0$: initial isotropic equivalent kinetic energy of the jet.\\
        $\theta_j$: half-opening angle of the jet.\\
        $\gamma_{\rm collision}$: Lorentz factor of the impulsive shell when energy injection begins, the collision point for the two shells.\\
        $\zeta$: the factor by which energy has increased at the end of energy injection.\\
        $s$: power law index for the mass added to the impulsive shell due to the collision at a given Lorentz factor.\\
        $n_0$: particle number density in the surrounding, unshocked medium.\\
        $p$: shock accelerated electron power-law index.\\
        $\varepsilon_e$: the fraction of the energy in the accelerated electrons.\\
        $\varepsilon_B$: the fraction of energy in the shock magnetic field.\\
        $\Gamma_0$: initial (pre-deceleration) bulk Lorentz factor of the impulsive shell.\\
        $\Xi_N$: the fraction of shocked electrons that participate in synchrotron emission.\\
        All priors are uniform within the range shown. The viewing angle is fixed at $\theta_{\rm v} = 0$, and redshift fixed at $z = 0.257$, the environment is assumed to be uniform, and the jet break given by geometric effects only. Posterior values indicate the distribution median and the 16th and 84th percentile.}
    \label{tab:priors}
\end{table}

\begin{figure*}
    \centering
    \includegraphics[width=\textwidth]{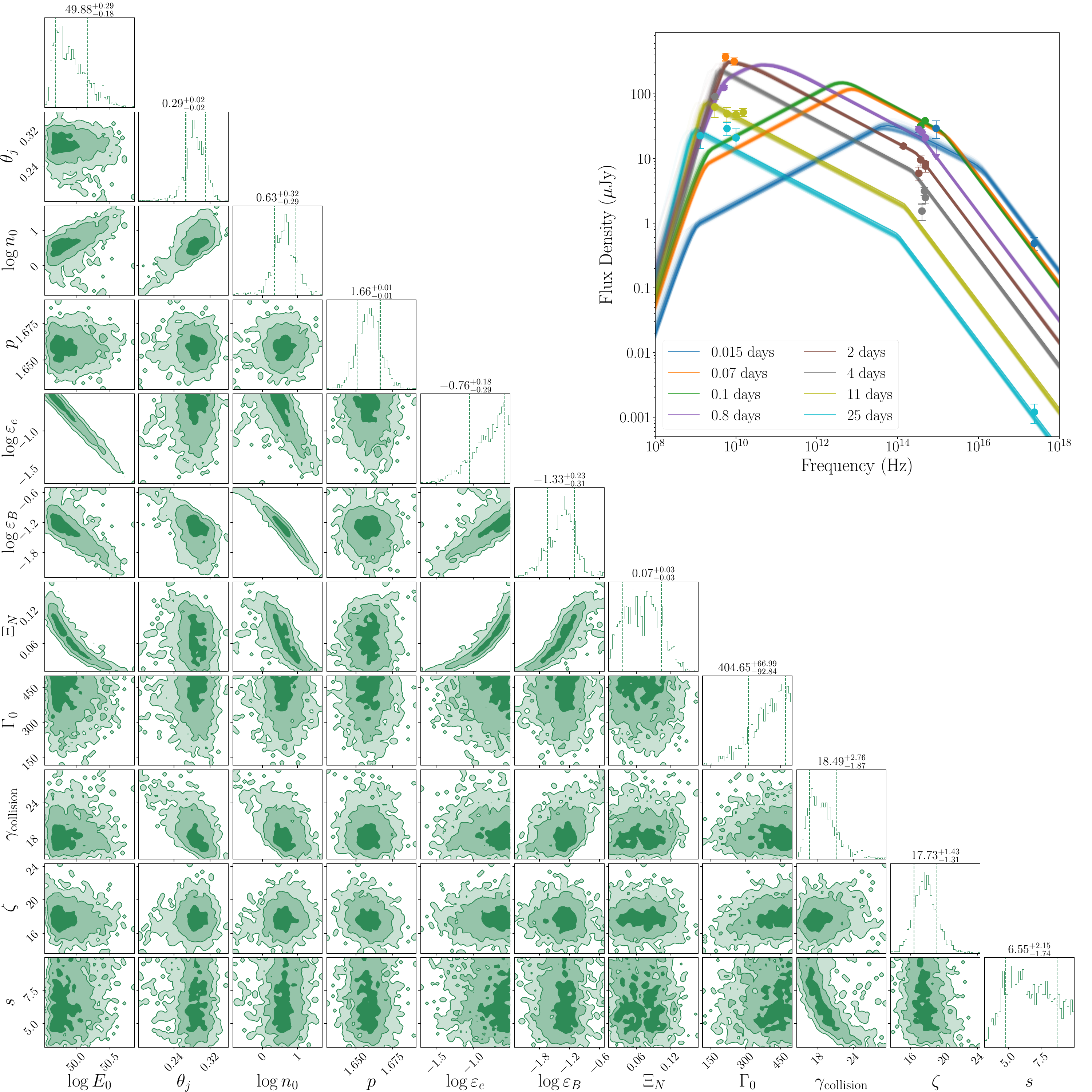}
    \caption{The model parameter posterior sample distribution found via {\sc Nessai} using a {\sc Redback} refreshed shock afterglow model, and the absorption/extinction corrected data shown in Figure~\ref{fig:powerlaws}. {\bf Top right}: spectral energy distribution for 200 randomly selected posterior sample outputs at given times vs the data. Note the radio excess at 0.07 days.}
    \label{fig:corner_AG}
\end{figure*}

The posterior distribution from this model fit is shown in Figure\,\ref{fig:corner_AG}. Some strong correlations are revealed, particularly between parameters that contribute primarily to the synchrotron emission, including the fraction of energy in the shocked accelerated electrons ($\varepsilon_e$) and the shock magnetic field ($\varepsilon_B$), the fraction of shocked electrons ($\Xi_N$), the initial isotropic equivalent kinetic energy ($E_0$), and the density of the unshocked CBM ($n_0$). 
We also see a correlation between the collision Lorentz factor, $\gamma_{\rm collision}$, and the power law index of the energy injection, $s$. 
This can be expected as the energy injection is $E\propto \gamma^{1-s}$ and the energy in the blastwave directly influences the power of the synchrotron emission, so a more rapid energy injection would require a lower collision Lorentz factor (i.e., a later onset of energy injection). 
The parameters that principally determine the impulsive blastwave dynamics, (e.g., $\Gamma_0$, $n_0$, $E_0$, $\zeta \times E_0$) have no obvious correlated degeneracies.
Additionally, the parameters that determine the spectral and temporal slopes of the synchrotron emission ($p$ and $\theta_j$) do not exhibit a correlation given that they are determined via the observed spectral energy distribution and the observed temporal behaviour (and similar to the blastwave dynamics, have little ambiguity).

The top right insert in Figure~\ref{fig:corner_AG} displays a selection of randomly drawn posterior sample spectral energy densities at fixed times compared to the data.
The spectral evolution is in good agreement between the data and the model, except for the radio band at GHz frequencies at 0.07 days (orange line vs points), where the model under-predicts this early radio data.
The radio data is in better agreement with the model from $\gtrsim 0.8$\,days, suggesting that the energy injection model fails to recreate the radio emission at, and throughout, the nominal energy injection period.
However, following the energy injection time frame, the resultant emission from this model is largely consistent with the observations.
Note that we do not include scintillation effects within our model, which can add additional uncertainty to the intrinsic flux density, especially at early times.
Nevertheless, even the predicted modulation index of 0.7 at 9\,GHz (see Table~\ref{tab:scint_pred}) cannot account for an order of magnitude discrepancy between the observations and model, supporting the existence of an additional emission component.  

\subsection{The violent collision}\label{sec:violent_model}

We now explore a potential source of the energy injection and the additional emission component detected at radio frequencies $<1$\,day post-burst.
Energy injection due to the collision of two relativistic shells generally follows one of two scenarios: the shell collision is mild/radiative, or the collision is violent/hot \citep{kumar2000, zhang2002}.
In the mild case, energy is deposited into the leading, or impulsive shell, increasing the initial blast-wave energy without establishing any new shocks. This case is also true if the jet is a Poynting-flux-dominated flow.
In the violent case, the shell collision results in strong supersonic shocks.
The condition for such shocks depends on the ratio of the energy of the catching, or injection shell to that of the impulsive shell, $E_i/E_j$, and the supersonic limit, given by $(\gamma_i/\gamma_j + \gamma_j/\gamma_i)/2 = \gamma_{ij} > 1.22$, where $\gamma_i$ and $\gamma_j$ are the Lorentz factors of the impulsive and the injection shell at the collision, respectively.
A strong shock is established for a supersonic collision where the energy ratio is above a threshold that depends on the radius of the collision and the spreading radius of the injection shell \citep[see Fig. 3 of ][]{zhang2002}.
The violent collision condition is therefore dependent on the shell energy ratio, the `thickness' of the injection shell ($\Delta_{\rm inj}$) compared to the collision radius ($\gamma_i^2 \Delta_{\rm inj}/R_{\rm col}$, where $\gamma^2 \Delta$ is $\sim$ the spreading radius), and  the velocity of the injection shell for an observer co-moving with the impulsive shell ($\gamma_{ij}$).

Using the parameter sample from the posterior distribution shown in Fig.\,\ref{fig:corner_AG}, we can determine if a strong shock is generated and estimate any additional flux due to the violent collision stage.
The refreshed shock model fit indicates that the total energy, once injection has ended, is $\zeta \sim18\times$ the initial impulsive shell energy -- as we follow \cite{zhang2002} for our collision model, our energy ratio being significantly above unity requires us to carefully consider the energy conservation equations during the violent collision stage. 
Additionally, in our case, the Lorentz factor of the injection shell is not constant.
Our model for the dynamics and electromagnetic emission from the violent collision is detailed in Appendix\,\ref{ap:model}.

There are two additional shocks established while the violent shock condition holds:
The first is a reverse shock travelling back into the injection shell, which is analogous to the reverse shock at deceleration for the impulsive shell;
the second is a forward shock that crosses the impulsive shell.
As this is a shock within already hot (relativistic) material, we must consider shock conditions that account for the additional internal energy and pressure.
The injection shell has more energy than the impulsive shell, and whereas the thin shell approximation well describes the impulsive shell, the injection shell is thick -- shell thickness can be approximated via a parameter, $\xi \equiv \left(t_{\rm dec}[1+z]/t_{90}\right)^{1/2}$, where $t_{\rm dec}$ is the deceleration time of the impulsive shell such that $t_{\rm dec} \propto E_0^{1/3} n^{-1/3} \Gamma_0^{-8/3}$, and is on the order of seconds for our model.
Where $\xi > 1$ the shell is considered thin, and for $\xi < 1$ the shell is thick.
We find $\xi \sim 1.6$ for the central parameters of our model posterior distribution, indicating a thin impulsive shell.
For the injection shell, the thick condition is met if the spreading radius is greater than the collision radius, $R_s > R_{\rm col}$, where $R_s \sim \gamma_{\rm collison}^2 \Delta_{\rm inj}$ and $R_{\rm col} \sim R_{\rm dec} (\gamma_{\rm collison} / \Gamma_0)^{-2/3}$.
Here $\Delta_{\rm inj} \sim c \delta t_{\rm inj}$, and $R_{\rm dec} = 2 \Gamma_0^2 t_{\rm dec} c$ is the deceleration radius.
Given the injection timescale of $\sim 1$\,day and a deceleration timescale on the order of seconds, the spreading radius is at least an order of magnitude larger than the collision radius, and indicates a thick injection shell.

The Lorentz factor in our injection shell is not constant but described as $M(>\Gamma) \propto \Gamma^{-s}$, where $M$ is the corresponding total mass of the shell faster than a given Lorentz factor.
The injection shell's Lorentz factor\footnote{From here we adopt the labelling index from \citet[][see their Figure 2]{zhang2002} for the various shell zones: A subscript `1' indicates the unshocked external medium, `2' is the shocked swept-up external medium (forward shock), `3' is the hot impulsive shell, `4' is the collision shocked hot impulsive shell, `5' is the shocked injection shell (reverse shock), and `6' is the unshocked injection shell. See also Figure~\ref{fig:collision_schem} for our setup.} at contact is then, $\gamma_6 \propto R^{-3/(1 + s)}$.
The leading zone of the impulsive shell, not shocked by the collision, will continue to decelerate as, $\gamma_3 \propto R^{-3/2}$.
To determine the instantaneous ratio of the energy density $e$ between the injection shocked impulsive shell material and the unaffected impulsive shell, plus the Lorentz factor of the injection shocked impulsive shell we must solve,
\begin{equation}
    \frac{e_4}{e_3} = \frac{n_6}{n_1}\frac{\left(\gamma_{46}-1\right)\left(4\gamma_{46} + 3\right)}{\left(\gamma_2 - 1\right)\left(4\gamma_2 + 3\right)},
    \label{eq:eratio}
\end{equation}
where $e_4 = e_5$, $\gamma_4 = \gamma_5$, then $\gamma_{46} = (\gamma_4/\gamma_6 + \gamma_6/\gamma_4)/2$, and the term $(4\gamma + 3)$ assumes an adiabatic index, $\hat\gamma = 4/3$. 
The general solution is given by $(\hat\gamma \gamma + 1)/(\hat\gamma - 1)$.
For the solution to eq.\,\ref{eq:eratio}, we need the Lorentz factor of the shocked impulsive shell, $\gamma_4$.
Given the conditions, $e_4/e_3 > 1$, and $\gamma_3 = \gamma_2$, in addition to knowing $\gamma_6$, $\gamma_2$, $n_1$, and $n_6$ we can solve the equality,
\begin{equation}
    \left(\frac{\gamma_4}{\gamma_3} + \frac{\gamma_3}{\gamma_4}\right)^2 = \frac{\left(1 + 3 e_4/e_3\right)\left(3 + e_4/e_3\right)}{4 e_4/e_3},
    \label{eq:gamma34}
\end{equation}
for the unknown $\gamma_4$.
As the shock jump conditions impose $\gamma_4 = \gamma_5$, we can determine the emission from the forward and reverse shocks in the injection -- impulsive shell collision.

\begin{figure}
    \centering\includegraphics[width=\columnwidth]{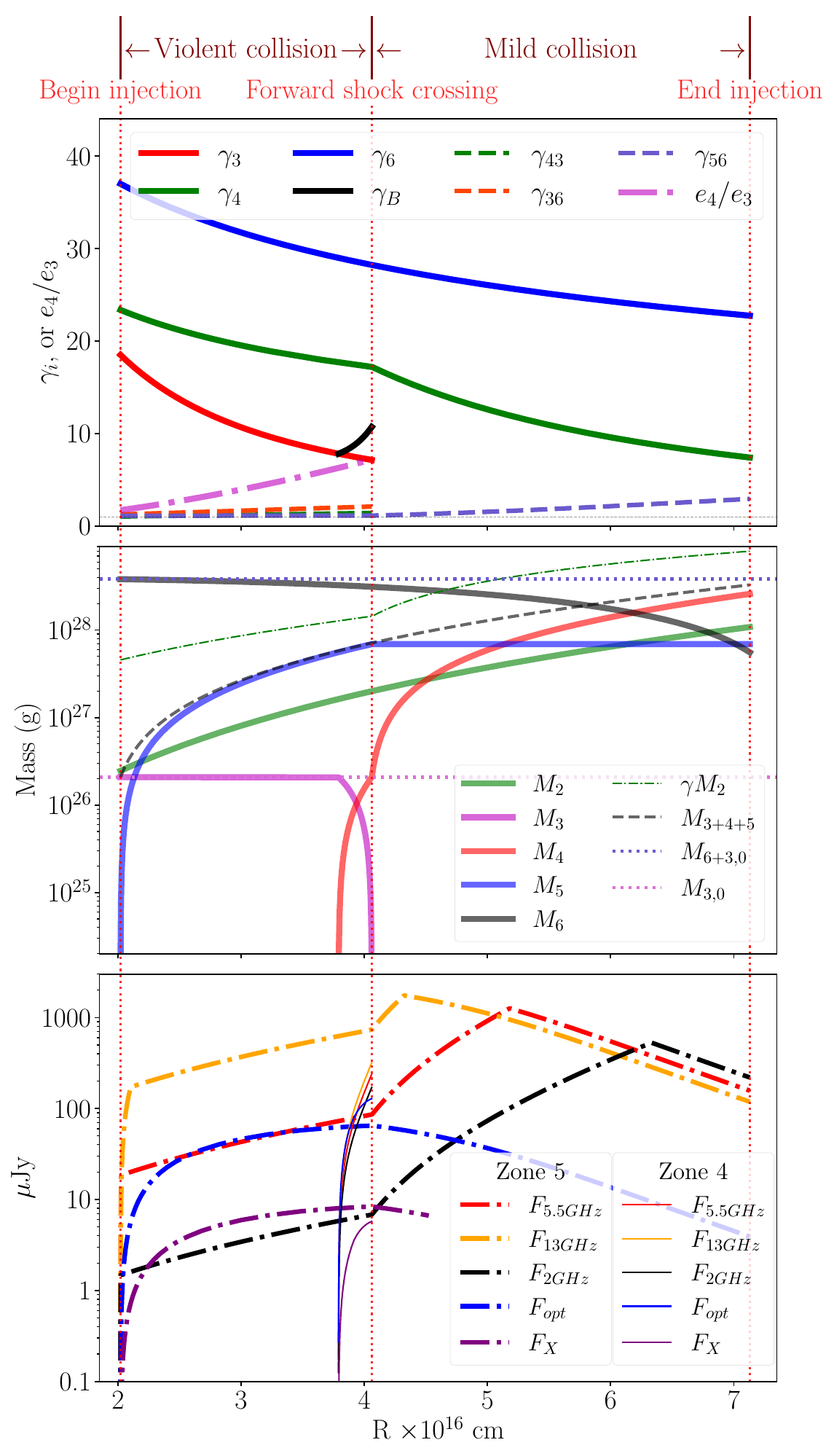}
    \caption{The required dynamical parameters from the refreshed shock model fit to the data indicate that the shell collision required by the model fit may have been `violent'. Solving the shock jump conditions and the requirement for a supersonic shock system, we can determine the duration of any violent collision and estimate the additional afterglow emission due to a reverse and forward shock system within the colliding shells.
    We adopt the labelling index from \citet[][see their Figure 2]{zhang2002} for the various shell zones: A subscript `1' indicates the unshocked external medium, `2' is the shocked swept-up external medium (forward shock), `3' is the hot impulsive shell, `4' is the collision shocked hot impulsive shell, `5' is the shocked injection shell (reverse shock), and `6' is the unshocked injection shell. See also Figure~\ref{fig:collision_schem} for our setup, and equation\,\ref{eq:gammaB} for the definition od $\gamma_b$.
    }
    \label{fig:collisions}
\end{figure}

The Lorentz factors, and shell masses with radius, as well as the synchrotron flux from the forward-reverse shock system are shown in Figure\,\ref{fig:collisions}.
The masses for each zone ensure mass, energy and momentum conservation for the dynamical system -- note that we do not consider radiative losses in our model.
See Appendix\,\ref{ap:model} for details.

\begin{figure*}
    \centering
    \includegraphics[width=\textwidth]{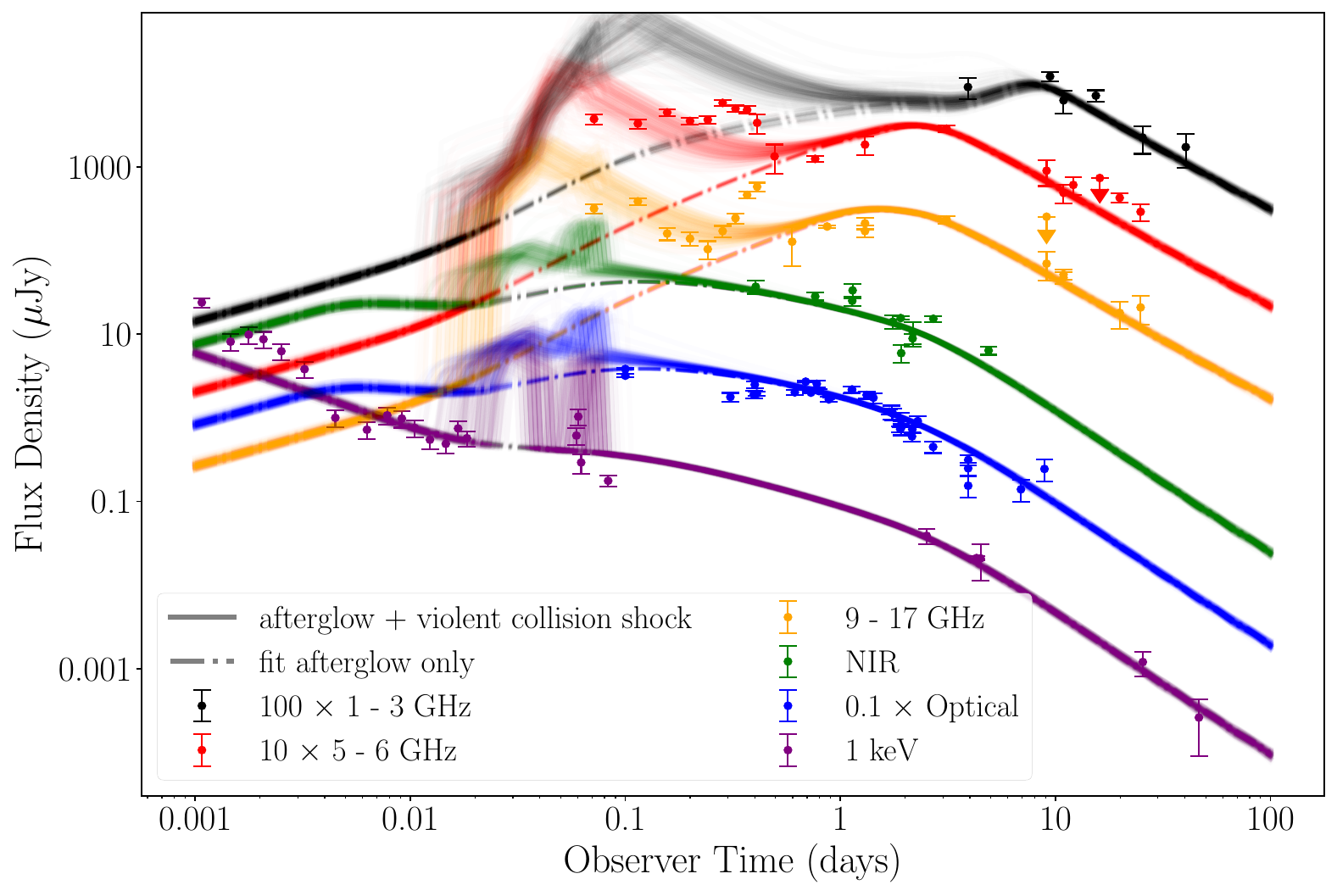}
    \caption{A randomly drawn sample of afterglow lightcurves from the {\sc Nessai} posterior samples, with the expected additional reverse/forward shock emission due to the violent collision conditions -- note that this extra emission component is not fit to the data, but provides early radio predictions that are more consistent with the data than the refreshed shock model alone. Note that the 5--6 GHz radio observations at $<0.5$\,days remain in excess of this model.
    However, parameters were fixed to those derived via the refreshed shock-only fit and may not be representative of those associated with the violent shell collision.  
    The additional emission component demonstrates that this violent collision phase, even without a detailed fit, can recreate the observed phenomenology of the early radio afterglow. For clarity, the model afterglow lightcurve at each `band' plotted within this figure assumes a single frequency for the emission, with the corresponding data covering the broader range indicated by the legend. The fit was performed using discrete model frequencies consistent with the observing frequency for each individual observation.}
    \label{fig:lightcurve_modelling__single}
\end{figure*}

In Figure\,\ref{fig:lightcurve_modelling__single}, we show how adding the flux from the forward and reverse shock system, given a violent collision at energy injection, can account for the model's under-prediction of the radio observations at $<0.5$\,days.
Note that we do not fit the violent collision parameters in the model, as this would result in too many free parameters.  
Instead, we feed in the posterior sample values (the distribution shown in Figure\,\ref{fig:corner_AG}) to solve this extra component -- this can be seen in the large spread in lightcurves for the violent collision emission vs the refreshed shock model.
For this figure we have added a magnetic amplification parameter $(\varepsilon_{B,5}/\varepsilon_B)^{1/2} \equiv R_B = 2$ \citep[see, ][]{harrison2013} to the model for the reverse shock emission, this ensures that the amplitude of the model flux density is consistent with that of the observations between $0.1 - 0.5$\,days.

From Figures~\ref{fig:collisions} and \ref{fig:lightcurve_modelling__single}, we can see that, because of the large imbalance in energy and thickness between the impulsive and the injection shells, that the forward shock crosses the impulsive shell before the end of the injection timescale.
We find that once this shock has crossed the impulsive shell, then the strong shock conditions are no longer satisfied and region `3' is replaced by region `4'.
At this time, the injection becomes mild and energy is gained by the impulsive shell without inducing any new emission sites.
No further energy is added to the reverse shock in the injection shell, and this shocked mass begins to relax.
As injection continues, the newly added mass is gained by region `4', without a shock. 
Region `4' acts like region `3' under mild energy injection, and the Lorentz factor declines as $\gamma_4 \propto R^{-3/2}$.
These dynamical effects are seen in the lightcurve as an early rise at collision, followed by a mildly increasing flux density.
As the collisional forward shock crosses region `3', we see a bright flare.
Interestingly, this flare coincides with that seen in the X-ray data. As we did not fit the violent collision model, this coincidence was unexpected, but may well help explain the observed flare.

The general agreement between the early radio data and the model reverse shock emission demonstrates that this violent collision phase, even without a detailed fit, can recreate the observed phenomenology of the early radio afterglow. 
Following the X-ray flare, the reverse shocked matter begins to relax and follows a mild decline.
Energy injection to the impulsive shell continues until $\sim 1$\,day, following the end of energy injection, the blastwave evolves as a classic decelerating forward shock system, with a geometric jet break at $\sim2$\,days.
 
As previously mentioned, we assume identical microphysical parameters $\left(\varepsilon_e,~\varepsilon_B,~\Xi_N,~p \right)$ for the violent collision reverse shock system as the initial refreshed shock model fit -- except for $\varepsilon_{B,5}$, for which we include the new parameter $R_B$.
Physically, there is no reason why the microphysical parameters should be identical, however, for simplicity, we have kept all but $\varepsilon_{B,5}$ the same to avoid excessive fine-tuning.
Additionally, if the collision Lorentz factor is reduced, the afterglow model at X-ray wavelengths does not over-predict the single data point at $\sim0.09$\,days, and additionally lowers the model optical flux density to be in better agreement at $0.1 - 0.5$\,days.
For the forward shock in the impulsive shell, we reduced the participation fraction by a factor $\Xi_{N,4} = 0.1\Xi{_N}$, which reduced the flare amplitude to be consistent with the observed flux density levels.

\section{Discussion}\label{sec:discussion}

\subsection{Radio SGRB comparisons}\label{sec:disc_radio_sgrb}

\begin{figure}
    \centering
    \includegraphics[width=\columnwidth]{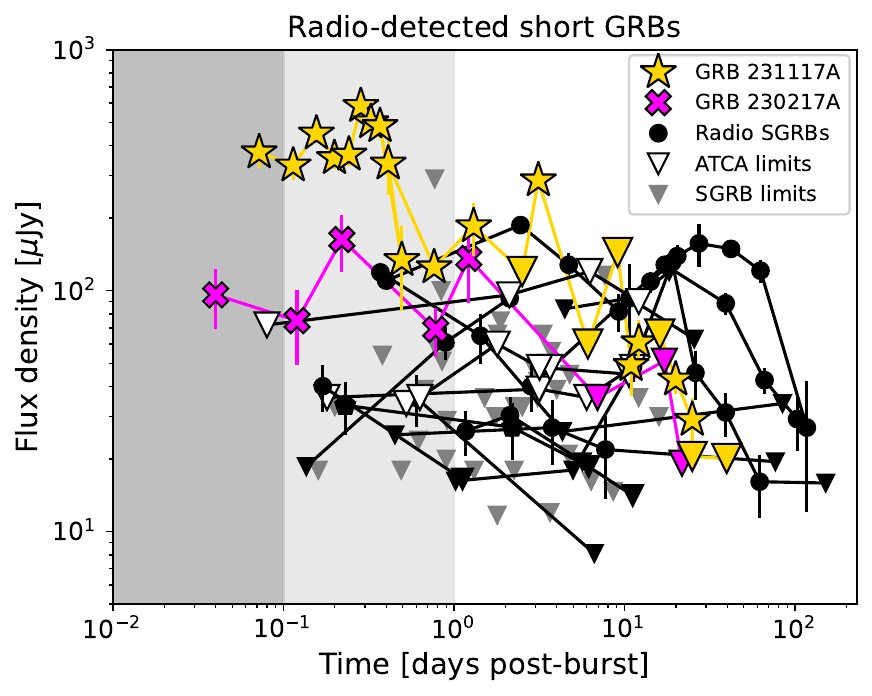}
    \caption{The light curves of SGRBs observed in the radio band between 5-8\,GHz. \thisgrb\ and GRB 230217A \citep{anderson24} are plotted as gold stars/triangles and magenta crosses/triangles, respectively, to illustrate the power of the ATCA rapid-response mode to probe $<0.1$\,days post-burst. The upper limits of the other SGRBs observed as part of this ATCA observing program are shown as white triangles \citep{anderson21,chastain24}. The other radio-detected SGRBs in the literature are plotted as black circles/triangles \citep{fong14,fong15,fong21,lamb19,troja19,laskar22,levan24,schroeder24,schroeder25}. 
    Radio $3\sigma$ upper limits on a population of SGRBs not detected in the radio band are plotted in grey \citep{fong15,schroeder25}. 
    All triangles represent $3\sigma$ upper limits on non-detections. 
    The timescales of $<0.1$ and $<1.0$\,day post-burst are shaded dark and light grey, respectively, to illustrate this undersampled parameter space in the radio band that we can now reach using ATCA rapid-response triggering.
    }
    \label{fig:radio_sgrbs}
\end{figure}

The early-time ($<1$\,day) radio sampling of the light curve of \thisgrb\ is unprecedented in GRB astrophysics.
Due to the proximity and brightness of the radio afterglow, we were able to split the ATCA rapid-response observations into 1-hr time bins, resulting in the first flux density measurement with a logarithmic central time of 1.7\,hr post-burst. This is the second earliest radio detection of a GRB in the published literature, with GRB 230217A being the current record holder at 1\,hr post-burst \citep{anderson24}. 
Figure~\ref{fig:radio_sgrbs} compares the $5-8$\,GHz radio light curves of all radio-detected SGRBs, highlighting the earliest detections (GRB 230217A and GRB 231117A) and limits $<0.1$\,days, a previously unexplored parameter space only now accessible due to the ATCA rapid-response mode \citep[][currently the only radio telescope operating this functionality at GHz frequencies]{anderson21}. 
Although optical and X-ray afterglow detections of \thisgrb\ were enough to infer energy injection, dense, early-time radio sampling has allowed us to establish the likely cause -- a violent collision between two shells launched simultaneously but with disparate Lorentz factors (see Sections~\ref{sec:violent_model} and further discussion in \ref{sec:disc_radio_origin}).

Of the 17 confirmed radio-detected SGRBs in the literature, multiwavelength modeling of six of these events attributes radio emission to a reverse shock, including \thisgrb. Of these six, three are likely classic reverse shocks associated with the jet interacting with the CBM \citep[GRB 160821B, GRB 200522A, GRB 230217A;][]{lamb19,fong21,anderson24}. 
However, the reverse shocks detected in \thisgrb\ as well as GRB 051221A and GRB 210726A may be associated with a violent shell collision \citep{soderberg06,schroeder24}, all three of which show evidence of energy injection into their afterglows \citep[as does GRB 160821B;][]{lamb19}. In five of these cases, early $<1$\,day radio observations were necessary to detect this emission component, the exception being GRB 210726A, which was undetected in the radio band until 11\,days, after which it underwent a radio flare \citep{schroeder24}.

\begin{figure}
    \centering
    \includegraphics[width=\columnwidth]{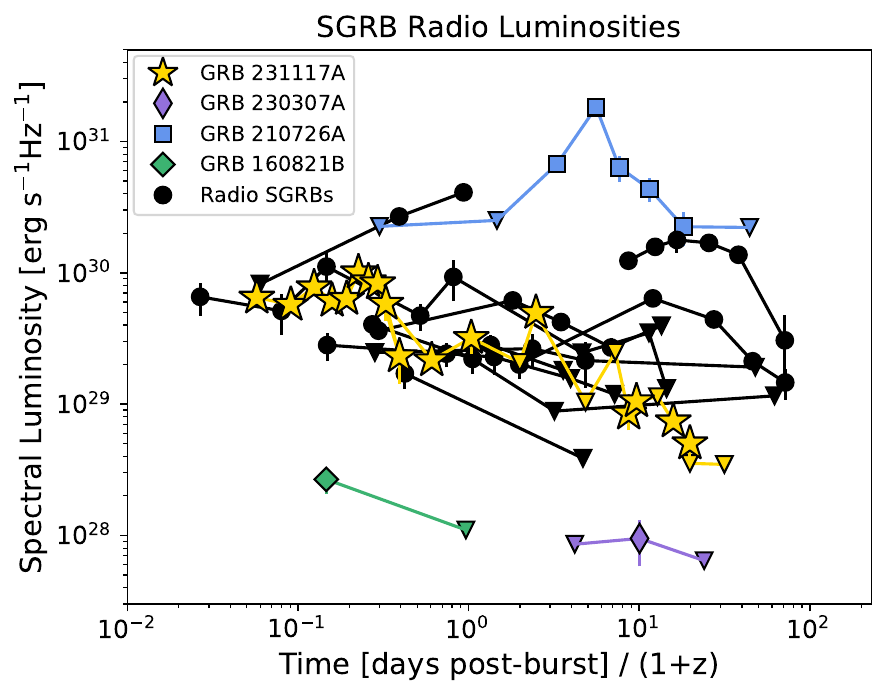}
    \caption{Luminosity comparisons of GRB 231117A (gold stars/triangles) and the other radio-detected SGRBs (black circles/triangles) between 5-8\,GHz \citep{fong14,fong15,fong21,laskar22,ferro23,levan24,schroeder25,anderson24}. The radio-detected short-duration GRB 210726A \citep[blue squares/triangles;][]{schroeder24} and GRB 160821B \citep[green large diamonds/triangles;][]{lamb19,troja19} that displayed energy injection (see Section~\ref{sec:disc_inject}) are also illustrated \citep[note that GRB 051221A was not detected in this frequency range;][]{soderberg06}.  
    We have also highlighted the long-duration merger GRB 230307A \citep[purple diamonds/triangles;][]{levan24} that had a long-lived central engine.
    }
    \label{fig:lum_sgrbs}
\end{figure}

We also compared the luminosity of \thisgrb\ with the larger radio-detected SGRB sample between 5-8\,GHz in Figure~\ref{fig:lum_sgrbs}. We highlight GRB 160821B and GRB 210726A as events that show evidence of energy injection \citep[note GRB 051221A is not plotted as it was not detected in this frequency range but has a similar luminosity to \thisgrb\ between 8-10\,GHz;][]{soderberg06}. We also draw attention to the long-duration GRB 230307A, which has an associated kilonova, indicating a BNS merger with a long-lived central engine \citep[][]{levan24}. The radio luminosity of \thisgrb\ is consistent with the majority of the radio-detected SGRB population. Interestingly, events with energy injection are at extremes, with GRB 210726A being the clear standout as it is an order of magnitude more luminous than \thisgrb. Meanwhile, GRB 160821B and GRB 230307A are at least an order of magnitude fainter. 
Although \thisgrb\ may not be unique in luminosity space, energy injection and/or long-lived central engines may represent some of the outliers.  

\subsection{Modelling comparisons}

Our preferred models can be compared to that of \citet{schroeder25} and with the larger GRB population. \citet{schroeder25} similarly identified a period of energy injection (a refreshed shock) in the \thisgrb\ afterglow that ended at $\sim1$\,day and also noted that the early radio data appears in excess of their model.
However, our refreshed shock afterglow model provides a better fit to both X-ray and optical data at $< 1$\,day -- noting that we have more data in our sample, which should improve the model fit.
When comparing the radio data fits to our refreshed shock model and that used in \cite{schroeder25}, both do a reasonable job for data $>0.5$\,days.
However, when a reverse shock is introduced, as Model 2 of \cite{schroeder25}, the radio data at 10 GHz is under-predicted.
The $9 - 17$\,GHz refreshed shock plus violent reverse shock of our model is in good agreement with the data, although radio variability at $0.2-0.3$\,days is not reproduced -- such variability indicates additional complexity within the system that is not included in our model.\footnote{Scintillation affects the 5--6 GHz and 9--17 GHz frequency ranges at this time. The observed chromatic variability between the radio bands could be due to ISS as explored in Section~\ref{sec:scint}.}
Our full model vs data at 5--6\,GHz under-predicts the $0.3 -0.5$\,day lightcurve by a factor of $\sim2 - 3$.
However, the light curve drops by this amount at $>0.5$,days and is largely in agreement with the model from this point on.
Although a full fit of our violent collision component should return better agreement between the model and data, overfitting becomes a significant risk when increasing the model complexity, given the limited data. 
As such, we do not attempt to fit the violent collision model in combination with the refreshed shock model but instead demonstrate that the data are consistent with such a scenario.

The refreshed shock model in \cite{schroeder25} assumed a delayed launch for the injection shell, whereas the assumption in our violent collision case was based on a common launch time but a slower Lorentz factor for the injection shell \citep[as in][]{zhang2002}.
In our scenario, the injection shell caught the impulsive shell when $\gamma_6 = 2\gamma_{\rm collison}$, where the collision Lorentz factor for the impulsive shell is given by the refreshed shock model fit to the data. 
This value approximately agrees with the last X-ray data point before the plateau used to infer injection at $\sim0.02$\,days (see Figures\,\ref{fig:powerlaws} and \ref{fig:lightcurve_modelling__single}).
We note that if the collision Lorentz factor is reduced in our model to give better agreement between the X-ray flux density at $\sim0.1$\,days, then the shell collision time is approximately coincident with that of the X-ray flare and just before the earliest radio data.

Analytically, \cite{schroeder25} also found an indication for $p<2$, although their fit value to the late afterglow gave $p = 2.24$ (their model invoked a uniform prior for $p$ from $2.01<p<2.99$).
With our increased data set, we have successfully joined all epochs consistently via the closure relations for, variously, a post-deceleration decline, a period of energy injection, and a final decline including a jet break.
Furthermore, our analytical estimate via closure relations of $p=1.7$ (see Section~\ref{sec:afterglow_model_1}) agrees with the model fit value of $p = 1.66\pm0.01$ (see Figure\,\ref{fig:corner_AG}).
Although a $p<2$ is not expected for shock-accelerated electrons in GRBs, there have been several observational cases that indicate $p<2$. 
For example, see Figure 12 in \citet{wang15}, the model fitting of \citet{aksulu22}, and the spectroscopy of GRB 221009A by \citet[][however, note this requires an early jet break, which was not observed in the long-term radio follow-up by \citealt{rhodes24}]{Levan2023_GRB221009A}.
The simplified single power-law assumption for shock-accelerated electrons ignores the contributions of thermal electrons and multiplicity within the shock zones. This may result in an apparent mismatch between the expected index and the model fit. Thus, we consider $p\sim 1.66$ from our model fits to be indicative of the average accelerated electron population distribution index $p$ for the afterglow to GRB\,231117A.

In general, those parameters derived for \thisgrb\ from the refreshed forward shock modeling (see Section~\ref{sec:main_model}) are consistent with the short and long GRB populations. Specifically, our value of $\varepsilon_e=0.17^{+0.09}_{-0.08}$ is consistent with the narrow distribution (centered at $\varepsilon_e=0.15$ for a homogeneous medium) derived by \citet[][]{beniamini17} using GRB radio light curve peaks.
It is also consistent with the range $0.1<\varepsilon_e<1.0$ derived from multi-wavelength modeling of a sample of 26 GRBs (4 short and 22 long) that have well sampled radio, optical and X-ray afterglows \citep{aksulu22}. 
Our derived value of $\varepsilon_B$ is also consistent with the large range derived by \citet{aksulu22}. 
Meanwhile, our derived value for $n_0=4.3^{+4.6}_{-2.1}$\,cm$^{-3}$ is among the highest for SGRBs \citep[][see their figure 9]{schroeder25}. However, \citet{schroeder25} showed that the radio-detected SGRB population was biased toward higher circumburst densities.

\subsection{Energy injection in SGRBs}\label{sec:disc_inject}

Afterglow modelling that includes energy injection in the form of refreshed shocks describe several GRB afterglows \citep[e.g.,][]{granot03}, including radio-detected short-duration events GRB 051221A, GRB 160821B, GRB 210726A \citep{soderberg06,lamb19,schroeder24}.
The energy injection required for both GRB\,160821B \citep[$\zeta\sim12.5$;][]{lamb19} and GRB\,231117A ($\zeta\sim18$) is more than an order of magnitude larger than the initial impulsive shell energy.
These suggest longer-lived central engine activity is possible for SGRB-like systems \citep[see e.g.][]{Gompertz15}, which is also demonstrated by the recent discovery of kilonova signatures in the afterglows of long-duration GRB\,211211A and GRB\,230307A \citep{rastinejad22,troja22nat,yang22nat,levan24,yang24nat} and the long-lasting extended emission of GRBs 050724, 060614 and 080503 \citep{gehrels06, gorosabel2006, perley2009}, showing that BNS mergers can successfully power prompt gamma-ray jets lasting $>2$\,s \citep[see also][]{Gompertz23}.
For GRB\,211211A, an analysis of the kilonova components (early blue and late red) revealed an inconsistency between the co-existence of both components within the observational constraints of this system. 
However, \cite{hamidani2024} demonstrated this tension could be resolved by allowing a longer-lived jet that results in late jet-ejecta interaction that produces the early blue emission misidentified as a blue kilonova component.
Such long-lasting jet activity may be analogous to the late energy injection scenario of the short GRBs 160821B and 231117A, where differences in system properties could result in long-duration merger-GRBs or late energy injection in SGRBs -- a study of the accretion properties of merger GRBs, long and short, is given by \cite{gottlieb2023}.

Long-lived central engines are also strongly supported by short-duration GRB 210726A, which displayed a late-time radio flare peaking at 19\,days post-burst. \citet{schroeder24} suggested the flare could be caused by energy injection, whether from a stratified jet with a distribution of Lorentz factors or via braking radiation from a magnetar remnant. This flare can also be described by the reverse shock from a violent shell collision similar to our scenario for \thisgrb\ but much more delayed, occurring at $\gtrsim11.2$\,days post-burst. All scenarios require a long-lived central engine, with the X-ray rebrightening suggesting a lifetime of $\sim1.4$\,days.

\subsection{Energetics and efficiency}\label{sec:disc_energ}

The initial impulsive shell in our model has an equivalent isotropic kinetic energy of $E_{\rm {Ki,iso}} = 7.6^{+7.2}_{-2.6} \times10^{49}$\,erg (equivalent to $E_0$ in our model, see Table~\ref{tab:priors}) and a final, refreshed energy $E_{\rm{Kf,iso}} = \zeta E_{\rm{Ki,iso}} =  1.35^{+1.28}_{-0.47}\times 10^{51}$\,erg. Using the reported spectral fit results from Konus-Wind we calculate, using a cosmological $k$-correction factor \citep{Bloom01}, a bolometric $E_{\gamma,{\rm iso}} \sim 1.3 \times 10^{51}$\,erg. 
The isotropic equivalent energetics of the final system are consistent with those found for SGRBs \citep{fong15, pandey2019}.
This indicates a $\gamma$-ray efficiency for the prompt emission of $\eta \sim 0.95$ if the GRB is entirely energy dissipated within the impulsive shell, or $\eta \sim 0.49$ if the GRB is a fraction of the total jet energy.
Given the model parameter uncertainties from the posterior distribution for the total kinetic energy, the latter of these is consistent with the expectation for SGRBs, $0.40 \lesssim \eta \lesssim 0.56$ \citep{fong15} -- this may indicate that the GRB emission is dominated by dissipation within the slower and radially stratified injection shell.

The posterior distribution for the jet's opening angle is $\theta_{\rm{j}} = 0.29 \pm 0.02 \mathrm{~rad} \equiv 16\mathring{.}6\pm 1\mathring{.}1$, where our model has assumed a top-hat jet and an on-axis observer (viewing angle $\theta_{\rm v}=0$), which is larger than $\sim10\mathring{.}4$ derived by \citet{schroeder25}.
This is consistent with the opening angles derived for the radio emitting SGRB population \citep[$\theta_{\rm{j}} \approx 3^{\circ} - 34^{\circ}$;][]{schroeder25}, 70\% of which have $\theta_{\rm{j}} >10^{\circ}$. 
However, assuming a more realistic geometry\footnote{As the typical line-of-sight angle, $\iota$, for a GRB is not on the central axis, the opening angle inferred via the jet break time that assumes $\iota = 0.0$ is a factor $\sim 3/2$ larger \citep{ryan15}.}, the true opening angle for the jet would be $\theta_{\rm{j}} \sim 9\mathring{.}5$ (the following collimation-corrected energies within parentheses assume this narrower jet).
The collimation-corrected jet energy is then, $E_{\rm{Ki}} \sim 3.2(0.9) \times 10^{48}$\,erg for the impulsive shell, and $E_{\rm{Kf}} \sim 5.7(1.7)\times10^{49}$\,erg for the total (final) kinetic energy.
This final collimation-corrected energy for the jet is consistent with that for the population of short duration GRBs \citep[$E_{\rm{K}} \sim 1.6^{+3.9}_{-1.3} \times10^{50}$\,erg;][]{fong15}.

A direct comparison can be made with GRB\,160821B in the refreshed shock model \citep{lamb19}.
The isotropic equivalent energies for GRB\,160821B were found to be, $E_{\rm{Ki,iso}} \sim 1.0\times 10^{50}$\,erg, and $E_{\rm{Kf,iso}} = 1.3\times10^{51}$\,erg -- a factor $\sim 12.5$ increase.
The jet opening angle for GRB\,160821B is much narrower than GRB\,231117A, at $\theta_{\rm{j}} \sim 1\mathring{.}9$ resulting in collimation-corrected energies of $E_{\rm{Ki}} \sim 5.5 \times 10^{46}$\,erg and $E_{\rm{Kf}} \sim 7.1\times10^{47}$ erg, an order of magnitude lower in jet energy than GRB\,231117A.
However, as the afterglow to GRB\,160821B indicates a very narrow jet, there could be significant jet energy carried in a wider structured component that is not apparent to an on-axis observer.

The energy injection for both GRB\,160821B and 231117A is of a similar order.
For GRB\,231117A, the injected energy is marginally larger than GRB\,160821B and the collision Lorentz-factor higher (the injection shell for GRB\,231117A had a minimum leading Lorentz-factor of approximately twice that in GRB\,160821B) -- these conditions result in a violent shock for GRB\,231117A.
Both of these cases indicate that, at least in some SGRBs, the majority of the jet energy can be carried by slower components and result in variability within the early afterglow.

\subsection{The origin of the early radio emission}\label{sec:disc_radio_origin}

The bright plateau of radio emission at very early times $0.07 - 1$\,days, in excess of the extrapolation of the optical to X-ray spectral energy density and lightcurve is consistent with this radio emission being from a distinct component.
The early declining X-ray lightcurve is a good indicator that the jet decelerated at $<0.003$\,days, and in such a scenario, any reverse shock due to the shells deceleration would cross and peak at this time.
No radio observations are available this early, and so the presence of a classic reverse shock cannot be ruled out, however, a classic reverse shock would rapidly fade following deceleration and become unobservable at the time of the radio excess seen in GRB\,231117A.
As the late afterglow indicates a period of significant energy injection, it is safe to assume that any injection shell may have had the required properties to establish a shock system between the impulsive and the injection shells.
Our analysis shows that although the leading shell is `thin', the injection shell is best described as `thick'.
This creates a unique situation that has not been clearly modelled in the literature -- to approximate the behaviour of the shocks within such a system we consider the mass and energy conservation, with details given in the Appendix\,\ref{ap:model}.
The profile of the injection shell, with a velocity stratification as $M(>\Gamma)\propto \Gamma^{-s}$, provides information about potential mass loading, jet acceleration, and/or shell stratification within SGRB (and therefore likely neutron star merger) systems.

The early radio emission in GRB\,231117A is likely dominated by a reverse shock within the injection shell as a result of a violent collision as demonstrated in Figure~\ref{fig:lightcurve_modelling__single}. 
Our model predicts that the earlier radio emission would have shown a step-like increase in flux density at the time of the collision.
Any initial reverse shock due to the impulsive shell deceleration would have faded before the collision.
Identifying both features in the afterglow would require even more sensitive and rapid follow-up observations within minutes of the explosion at GHz frequencies for all nearby GRB afterglows.
Additionally, as the forward shock due to the violent collision crosses the impulsive `thin' shell, there is a distinctive flare in the model lightcurves.
For GRB\,231117A, this predicted flare is coincident with an observed flare at X-ray frequencies $\sim0.07$\,days post-burst.

\begin{figure}
    \centering
    \includegraphics[width=\columnwidth]{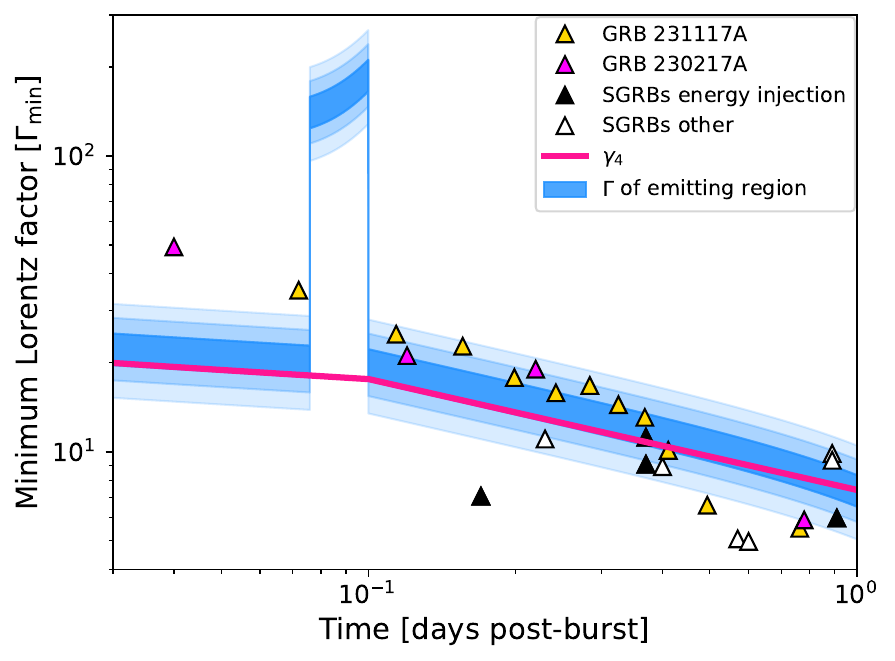}
    \caption{Comparison between the minimum Lorentz factor derived from the radio detections of GRB 231117A (gold triangles, see Section~\ref{sec:disc_radio_origin}) and the shell collision model outlined in Section~\ref{sec:violent_model} as a function of time. 
    The deep pink line shows the median model-derived Lorentz factor of the shocked impulsive shell ($\gamma_4$).
    The minimum Lorentz factor constraints from the radio observations up to $\sim1$\,day post-burst reflect the effective model-derived Lorentz factor of the emitting region ($\Gamma$), which is a combination of Lorentz factors including $\gamma_5 \gamma_{56}$ dominating the reverse shock, $\gamma_4 \gamma_{\rm{B}}$ for the shocked impulsive shell, and $\gamma_4$ from the forward shock. 
    This is shown by the blue shaded region, where each gradient in the colour indicates the $\sigma$ uncertainty in the model values -- note $\gamma_4 = \gamma_5$. 
    We also plot the short duration GRB 230217A, which is the current radio-derived minimum Lorentz factor record holder \citep[magenta triangles;][]{anderson24}. 
    All other SGRBs with radio detections $<1$\,day post-burst are plotted \citep[white triangles;][]{berger05,fong15,fong21,schroeder25}, with those that (may have) exhibited energy injection in their afterglows indicated by black triangles \citep{soderberg06,fong14,fong21,lamb19}. 
    } 
    \label{fig:gamma}
\end{figure}

The radio detections between $0.07 - 1$\,day post-burst can also be used to place a lower limit on the Lorentz factor ($\Gamma_{\mathrm{min}}$) of the injection shell, which is equivalent to $\gamma_5 \gamma_{56}$ defined in the violent collision model in Section~\ref{sec:violent_model}.
Assuming the emitting region has expanded to a maximum size of $ct$ then the brightness temperature ($T_{\rm{b}}$) for each flux density measurement can be calculated following equation 1 in \citet{anderson14}. 
The minimum Lorentz factor ($\Gamma_{\mathrm{min}}$) can be calculated according to $T_{\rm{b}}/T_{\rm{B}} \lesssim \Gamma^{3}$ \citep{galama99}, where $T_{\rm{B}}\sim10^{12}$\,K is the maximum brightness temperature from the inverse Compton limit. 
The minimum Lorentz factor for GRB 231117A as shown in Figure~\ref{fig:gamma} is calculated using the $1\sigma$ lower limit on the 5.5\,GHz flux density measurements in Table~\ref{tab:radio_flux}. 

We compare the minimum Lorentz factors to those of the model by overlaying the range in model-predicted $\gamma$ for the dominant emission region: reverse shock, collision shocked impulsive shell, forward shock -- depending on the time after burst.
This is plotted as a function of time in Figure~\ref{fig:gamma}\footnote{Note that as we assume both shells are ejected at the same time and that the impulsive shell `clears the path' for the injection shell, our model estimates are technically lower-limits and any delay in the injection shell launch or energy spent sweeping a path would require a higher Lorentz factor at collision for the injection shell.}.  
During energy injection, a close match exists between the Lorentz factor lower limits derived from the radio data and those of the dominant emission region within the energy injection model, further supporting the violent collision interpretation. 
Given the reverse shock dominates the radio emission at early times, the inferred minimum Lorentz factor should be equivalent to $\gamma_{5} \gamma_{56}$ in our model, with the exception of the rapid flare, seen in X-rays but predicted by our model at radio wavelengths at $\lesssim0.1$\,days.
The flare Lorentz factor is uncertain, but the model predicts a rapid and hot shock that crosses the impulsive shell with an effective emission Lorentz factor on the order $\gamma_4 \gamma_{\rm{B}} \sim 100$.
Energy injection continues until $\sim0.3$\,days, however, the violent shock condition is no longer met $\gtrsim0.1$\,days and the Lorentz factor declines as $\gamma \propto t^{-3/8}$ \citep{zhang2002}.
After the flare at $>0.1$\,days, the reverse shock stalls and with no new shocked electrons within the reverse shock, the hot electrons cool and relax until emission from the forward shock becomes dominant at $\sim 1$\,day.

\section{Conclusions}

In this paper we present an extensive multi-wavelength analysis of the short \thisgrb, performing broadband modeling of the afterglow from radio to X-ray wavelengths. This high time and spectral resolution of the early-time ($<1$\,day) radio coverage is unprecedented for any GRB, long or short, providing direct insight into the complex structure of their relativistic outflows and therefore a window into the activity of the central engine. Our analysis leads to the following conclusions relating to the blast wave underlying the afterglow of \thisgrb.

\begin{itemize}
    \item The ATCA rapid-response observation of \thisgrb\ at 5.5 and 9\,GHz, beginning 1.3\,hr post-burst was split into 1 hour time bins across the full 9 hour integration. 
    These measurements demonstrate that the afterglow had significant variability within this observation, likely caused by interstellar scintillation. We used this scintillation to place the earliest upper limit on a GRB source size to date of $<1\times10^{16}$\,cm (see Section~\ref{sec:scint} and Table~\ref{tab:scint_pred}).  
    \item When the ATCA data is combined with \textit{e}-MERLIN and published VLA data \citep{schroeder25}, the light curve plateaus up to $\sim1$\,day post-burst (see Figure~\ref{fig:lc_spec}), which was also observed at optical and X-ray wavelengths (see Figure~\ref{fig:powerlaws}). Such plateaus are indicative of energy injection.
    \item The multi-wavelength modeling combined radio, optical and X-ray observations of \thisgrb\ beginning within minutes and up to 40 days post-burst. We employ a refreshed forward shock model for \thisgrb, with a period of significant energy injection between $0.02-1$\,day post-burst. At $>1$\,day, the blast wave evolves as a classic forward shock and requires a low value of the accelerated electron distribution index, $p=1.66 \pm 0.01$. This afterglow model has a jet break at $\sim2$\,days and gives a jet half-opening angle of $\theta_{\rm{j}} =16\mathring{.}6 \pm 1\mathring{.}1$ and a final collimation-corrected jet energy of $E_{\rm{Kf}} \sim 5.7 \times 10^{49}$\,erg (for an on-axis, $\theta_{\rm v} = 0$, observer). 
    \item The refreshed forward shock model cannot account for the flaring radio emission observed during the energy injection phase (see Figure~\ref{fig:corner_AG}), indicating the need for an additional emission component. We suggest the additional component is due to a violent collision between two relativistic shells  \citep{zhang2002}. The impulsive (leading) shell in our model has a high Lorentz factor, and its passage clears a channel or path for the trailing injection shell. The injection shell has a higher energy and a stratified distribution of slower Lorentz factors (see Section~\ref{sec:violent_model}). The energy injection episode is the result of the injection shell catching the decelerating impulsive shell. For our parameters, the collision is above the supersonic limit (a violent collision) and two additional shocks are established; a reverse shock in the injection shell, and a forward shock that crosses the impulsive shell (see Figure~\ref{fig:collisions} and Appendix~\ref{ap:model}.)
    \item Assuming identical microphysical parameters $\left(\varepsilon_e,~\varepsilon_B,~\Xi_N,~p \right)$ for the entire system (except for $\varepsilon_{B,5}$ within the injection shell), we add the resulting flux from the violent collision induced shock system into our model in Figure~\ref{fig:lightcurve_modelling__single}. This additional flux component qualitatively accounts for the radio emission observed $<0.5$\,days post-burst, providing a robust explanation for this excess.
    This model also predicts a flare at X-ray that is incidentally coincident with an observed X-ray flare. This unintentional coincidence is additionally supportive of the violent shell collision scenario. 
    \item Radio detections of reverse shocks probe the lower limit on the minimum Lorentz factor of the blastwave \citep[e.g.][]{anderson14,anderson18,anderson24,bright23}. Figure~\ref{fig:gamma} demonstrates that the minimum Lorentz factors required by our model agree with the early-time ATCA detections of \thisgrb\ (see Section~\ref{sec:disc_radio_origin}), further supporting a violent shell collision scenario.  
\end{itemize}

Due to the ATCA rapid-response mode, our triggered observation of \thisgrb\ is the second-earliest published radio detection of a GRB to date. We therefore continue to demonstrate that such telescope functionalities allow us to probe the previously unexplored parameter space of $<0.01$\,days post-burst in the radio band (see Figure~\ref{fig:radio_sgrbs}). Without this rapid-response radio detection of \thisgrb, multi-wavelength modeling would not have been able to identify a violent shell collision as the likely source of energy injection, and thus provide direct insight into the activity of the central engine. Similar rapid follow-up observations of SGRBs will allow us to study the diversity of their central engines, particularly those events with extended or prolonged gamma-ray emission \citep[e.g.][]{rastinejad22,levan24}.
Indeed, triggering functionalities are planned for both SKA telescopes.\footnote{https://www.skao.int/en/resources/402/key-documents} Given that SKA-Mid is forecast to obtain a $3\sigma$ sensitivity of $\sim6\mu$\,Jy/beam on 1\,minute timescales in both the 4.6--8.5 and 8.3--15.4\,GHz bands (assuming an elevation of $45^{\circ}$ and a Robust weighting of 0),\footnote{https://sensitivity-calculator.skao.int/mid} such rapid-response observations will provide detections of more and fainter SGRBs, providing unprecedented detail of their early radio afterglows on minute timescales.


\section{Acknowledgments}

We acknowledge the Whadjuk Nyungar people as the traditional owners of the land on which the Bentley Curtin University campus is located, where the majority of this research was conducted. 

We thank the referee for their recommendations on this manuscript.
GEA is the recipient of an Australian Research Council Discovery Early Career Researcher Award (project number DE180100346) funded by the Australian Government.
GPL is supported by a Royal Society Dorothy Hodgkin Fellowship (grant Nos. DHF-R1-221175 and DHF-ERE-221005).
BPG acknowledges support from STFC grant No. ST/Y002253/1 and from The Leverhulme Trust grant No. RPG-2024-117.
LR acknowledges support from the Canada Excellence Research Chair in Transient Astrophysics (CERC-2022-00009). 
T-WC acknowledges the financial support from the Yushan Fellow Program by the Ministry of Education, Taiwan (MOE-111-YSFMS-0008-001-P1) and the National Science and Technology Council, Taiwan (NSTC 114-2112-M-008-021-MY3)
RLCS acknowledges support from The Leverhulme Trust grant No. RPG-2023-240.
SY and Z-NW acknowledge the funding from the National Natural Science Foundation of China under Grant No. 12303046 and the Henan Province High-Level Talent International Training Program.
AA acknowledges the support of the Breakthrough Listen project. Breakthrough Listen is managed by the Breakthrough Initiatives, sponsored by the Breakthrough Prize Foundation.
RB acknowledges funding from the Italian Space Agency, contract ASI/INAF n. I/004/11/6.
D acknowledges the support from STFC grant No. ST/Y002253/1.
NPMK acknowledges UK Space Agency support.
DBM is funded by the European Union (ERC, HEAVYMETAL, 101071865). Views and opinions expressed are, however, those of the authors only and do not necessarily reflect those of the European Union or the European Research Council. Neither the European Union nor the granting authority can be held responsible for them. The Cosmic Dawn Center (DAWN) is funded by the Danish National Research Foundation under grant DNRF140.
MN is supported by the European Research Council (ERC) under the European Union’s Horizon 2020 research and innovation programme (grant agreement No.~948381).
PTOB, NRT acknowledges support from STFC grant ST/W000857/1.
SJS acknowledges funding from STFC Grants ST/Y001605/1, ST/X006506/1, ST/T000198/1, a Royal Society Research Professorship and the Hintze Charitable Foundation. 
IW is supported by the UKRI Science and Technology Facilities Council (STFC).
This work was funded by ANID, Millennium Science Initiative, ICN12\_009.

This work was supported by software support resources awarded under the Astronomy Data and Computing Services (ADACS) Merit Allocation Program. ADACS is funded from the Astronomy National Collaborative Research Infrastructure Strategy (NCRIS) allocation provided by the Australian Government and managed by Astronomy Australia Limited (AAL).

We acknowledge the Gomeroi people as the traditional owners of the Australia Telescope Compact Array (ATCA) observatory site. 
The ATCA is part of the Australia Telescope National Facility, which is funded by the Australian Government for operation as a National Facility managed by CSIRO. 
\textit{e}-MERLIN is a National Facility operated by the University of Manchester at Jodrell Bank Observatory on behalf of STFC, part of UK Research and Innovation.
The \textit{e}-MERLIN observations obtained in this work were the first to make use of the new Lovell telescope frequency flexibility system. We thank Neil Roddis and the engineering team at Jodrell Bank for the commissioning and testing of the new system.
We thank the staff of the Mullard Radio Astronomy Observatory, University of Cambridge, for their support in the maintenance, and operation of AMI. We acknowledge support from the European Research Council under grant ERC-2012-StG-307215 LODESTONE.
The MeerKAT telescope is operated by the South African Radio Astronomy Observatory, which is a facility of the National Research Foundation, an agency of the Department of Science and Innovation.

This publication makes use of data collected at Lulin Observatory, partially supported by TAOvA under NSTC grant 112-2740-M-008-002.
The Pan-STARRS1 Surveys (PS1) and the PS1 public science archive have been made possible through contributions by the Institute for Astronomy, the University of Hawaii, the Pan-STARRS Project Office, the Max-Planck Society and its participating institutes, the Max Planck Institute for Astronomy, Heidelberg and the Max Planck Institute for Extraterrestrial Physics, Garching, The Johns Hopkins University, Durham University, the University of Edinburgh, the Queen's University Belfast, the Harvard-Smithsonian Center for Astrophysics, the Las Cumbres Observatory Global Telescope Network Incorporated, the National Central University of Taiwan, the Space Telescope Science Institute, the National Aeronautics and Space Administration under Grant No. NNX08AR22G issued through the Planetary Science Division of the NASA Science Mission Directorate, the National Science Foundation Grant No. AST-1238877, the University of Maryland, Eotvos Lorand University (ELTE), the Los Alamos National Laboratory, and the Gordon and Betty Moore Foundation.
Pan-STARRS is a project of the Institute for Astronomy of the University of Hawaii, and is supported by the NASA SSO Near Earth Observation Program under grants 80NSSC18K0971, NNX14AM74G, NNX12AR65G, NNX13AQ47G, NNX08AR22G, 80NSSC21K1572  and by the State of Hawaii.
The Gravitational-wave Optical Transient Observer (GOTO) project acknowledges support from the Science and Technology Facilities Council (STFC, grant numbers ST/T007184/1, ST/T003103/1, ST/T000406/1, ST/X001121/1 and ST/Z000165/1) and the GOTO consortium institutions; University of Warwick; Monash University; University of Sheffield; University of Leicester; Armagh Observatory \& Planetarium; the National Astronomical Research Institute of Thailand (NARIT); University of Manchester; Instituto de Astrofísica de Canarias (IAC); University of Portsmouth; University of Turku.
Based on observations collected at the European Organisation for Astronomical Research in the Southern Hemisphere under ESO programme 110.24CF (Stargate collaboration)"
Based on observations collected at the European Southern Observatory under ESO programme 112.25JQ and at the La Silla Paranal Observatory under programme IDs 1103.D-0328 and 106.216C.

This work made use of the Astro-COLIBRI platform \citep{reichherzer21}.
This research makes use of {\sc Astropy}, a community-developed core Python package for Astronomy \citep{astropy22},
{\sc numpy} \citep{vanderWalt_numpy_2011} and {\sc scipy} \citep{Jones_scipy_2001} python modules. This research also makes use of {\sc matplotlib} \citep{hunter07}. 
This research has made use of NASA's Astrophysics Data System. 
This research has made use of SAOImage DS9, developed by the Smithsonian Astrophysical Observatory.
This research uses {\sc Nessai} \citep{nessai}, and {\sc Redback} \citep{redback} for transient modeling and inference.
This work made use of data supplied by the UK Swift Science Data Centre at the University of Leicester.
This research has made use of the NASA/IPAC Extragalactic Database (NED), which is funded by the National Aeronautics and Space Administration and operated by the California Institute of Technology.


\appendix

\section{Observational datasets}\label{ap:data}

All the new optical and radio data that were used to perform the afterglow modeling in Section~\ref{sec:afterglow_model} are listed in Tables~\ref{tab:optical} and \ref{tab:radio_flux}, respectively. The host galaxy photometry is presented in Table~\ref{tab:phot:host}.

\begin{center}
\begin{longtable*}{llll}
\caption{The optical measurements of GRB 231117A. The time post-burst corresponds to the centre of the integration time. All errors are $1\sigma$ and all upper limits are $3\sigma$.}\\
\hline\hline \multicolumn{1}{l}{Telescope} &  \multicolumn{1}{l}{Time post-burst} & \multicolumn{1}{l}{Filter} & \multicolumn{1}{l}{Magnitude} \\
& \multicolumn{1}{l}{(days)} & & \multicolumn{1}{l}{ (AB)} \\
\hline
\endfirsthead

\hline\hline \multicolumn{1}{l}{Telescope} & \multicolumn{1}{l}{Time post-burst} & \multicolumn{1}{l}{Filter} & \multicolumn{1}{l}{Magnitude} \\
& \multicolumn{1}{l}{(days)} & & \multicolumn{1}{l}{ (AB)} \\
\hline
\endhead

\hline\hline
\endfoot

\hline\hline
\endlastfoot

UVOT & 0.002 & $white$ & $> 20.77$ \\
UVOT & 0.008 & $white$ & $> 20.15$ \\
UVOT & 0.011 & $u$ & $> 21.10$ \\
UVOT & 0.013 & $uvw2$ & $> 20.65$ \\
UVOT & 0.013 & $uvw2$ & $> 20.79$ \\
UVOT & 0.013 & $b$ & $> 19.00$ \\
UVOT & 0.013 & $v$ & $> 18.88$ \\
UVOT & 0.014 & $uvw1$ & $> 20.65$ \\
UVOT & 0.015 & $white$ & $20.23 \pm 0.33$ \\
UVOT & 0.019 & $white$ & $20.44 \pm 0.43$ \\
UVOT & 0.060 & $uvw2$ & $> 19.74$ \\
UVOT & 0.062 & $v$ & $> 18.03$ \\
UVOT & 0.064 & $uvm2$ & $> 19.80$ \\
UVOT & 0.067 & $uvw1$ & $> 20.48$ \\
UVOT & 0.140 & $u$ & $> 20.50$ \\
UVOT & 0.213 & $b$ & $> 17.70$ \\
UVOT & 0.789 & $white$ & $22.42 \pm 1.09$ \\
UVOT & 2.511 & $white$ & $23.44 \pm 1.13$ \\
UVOT & 3.700 & $white$ & $> 22.17$ \\
UVOT & 4.328 & $white$ & $> 22.20$ \\
UVOT & 5.750 & $white$ & $> 22.62$ \\
UVOT & 6.377 & $white$ & $> 22.30$ \\
UVOT & 7.162 & $white$ & $> 20.92$ \\
\hline
SLT & 0.308 & $r$ & $20.78 \pm 0.13$ \\
SLT & 1.325 & $r$ & $20.72 \pm 0.11$ \\
SLT & 1.425 & $i$ & $20.80 \pm 0.15$ \\
SLT & 1.453 & $g$ & $21.13 \pm 0.12$ \\
SLT & 1.490 & $z$ & $> 19.94$ \\
SLT & 2.291 & $i$ & $21.50 \pm 0.17$ \\
\hline
PS1 & 1.129 & $g$ & $20.74 \pm 0.16$ \\
PS1 & 1.131 & $r$ & $20.56 \pm 0.11$ \\
PS1 & 1.134 & $i$ & $20.55 \pm 0.09$  \\
PS1 & 1.136 & $z$ & $20.40 \pm 0.14$ \\
PS1 & 1.140 & $y$ & $20.09 \pm 0.21$  \\
PS1 & 2.146 & $g$ & $22.15 \pm 0.15$ \\
PS1 & 2.154 & $r$ & $21.96 \pm 0.14$ \\
PS1 & 2.161 & $i$ & $21.75 \pm 0.11$ \\
PS1 & 2.168 & $z$ & $21.53 \pm 0.16$ \\
PS1 & 2.176 & $y$ & $21.35 \pm 0.35$  \\
\hline
NTT & 0.884 & $R$ & $20.85 \pm 0.08$ \\
NTT & 6.916 & $R$ & $23.54 \pm 0.32$ \\
NTT & 8.897 & $R$ & $22.93 \pm 0.31$ \\
\hline
LT & 0.690 & $i$ & $20.32 \pm 0.03$ \\
LT & 0.730 & $r$ & $20.65 \pm 0.03$ \\
LT & 0.764 & $z$ & $20.27 \pm 0.12$ \\
LT & 0.777 & $i$ & $20.38 \pm 0.10$ \\
LT & 0.794 & $r$ & $20.60 \pm 0.06$ \\
LT & 1.690 & $i$ & $21.22 \pm 0.08$ \\
LT & 1.730 & $r$ & $21.23 \pm 0.21$ \\
LT & 1.753 & $z$ & $21.03 \pm 0.20$ \\
LT & 1.758 & $i$ & $21.28 \pm 0.171$ \\
LT & 1.763 & $r$ & $> 21.05$ \\
LT & 1.769 & $g$ & $> 21.09$ \\
LT & 2.700 & $i$ & $22.27 \pm 0.17$ \\
LT & 2.740 & $r$ & $> 21.737$ \\
\hline
LOT & 0.393 & $g$ & $20.69 \pm 0.1$ \\
LOT & 0.396 & $r$ & $20.70 \pm 0.11$ \\
LOT & 0.400 & $i$ & $20.41 \pm 0.11$ \\
LOT & 0.404 & $z$ & $19.98 \pm 0.19$ \\
LOT & 0.409 & $r$ & $20.68 \pm 0.07$ \\
LOT & 3.330 & $i$ & $> 21.14$ \\
\hline
VLT-FORS2 & 1.888 & $g$ & $21.84 \pm 0.09$ \\
VLT-FORS2 & 1.899 & $R$ & $21.74 \pm 0.09$ \\
VLT-FORS2 & 1.908 & $I$ & $21.45 \pm 0.21$ \\
VLT-FORS2 & 1.918 & $Z_{\rm GUNN}$ & $21.97 \pm 0.26$ \\
VLT-FORS2 & 3.932 & $R$ & $22.66 \pm 0.16$ \\
VLT-FORS2 & 8.955 & $Z_{\rm GUNN}$ & $> 21.70$ \\
\hline
VLT-HAWKI & 1.908 & $Ks$ & $20.92 \pm 0.05$ \\
VLT-HAWKI & 4.891 & $Ks$ & $21.89 \pm 0.12$ \\
VLT-HAWKI & 6.925 & $Ks$ & $> 23.83$ \\
VLT-HAWKI & 24.90 & $Ks$ & $> 20.74$ \\
\hline
GTC & 2.707 & $K_{\rm AB}$ &  $20.94 \pm 0.09$  \\
GTC & 8.726 & $z$ & $> 22.41$ \\
\hline
GOTO & 0.289 & $L$ & $> 19.70$ \\
GOTO & 0.684 & $L$ & $> 19.37$ \\
GOTO & 1.674 & $L$ & $> 18.98$ \\
GOTO & 1.694 & $L$ & $> 19.37$ \\
GOTO & 1.716 & $L$ & $> 18.98$ \\
GOTO & 2.702 & $L$ & $> 18.99$ \\
\hline
CAHA & 0.614 & $i$ & $20.64 \pm 0.07$ 
\label{tab:optical}
\end{longtable*}
\end{center}

\begin{table*}
    \centering
    \begin{tabular}{lllll}
    \hline
    \hline
    Telescope & Time post-burst & $\nu$ & Flux density & $3\sigma$ threshold \\
    & (days) & (GHz) & ($\mu$Jy/beam) & ($\mu$Jy/beam) \\
    \hline
    ATCA & 0.07 & 5.5 & $\bf{375 \pm 53}$ &  \\ 
    ATCA & 0.11 & 5.5 & $\bf{331 \pm 42}$ & \\ 
    ATCA & 0.16 & 5.5 & $\bf{449 \pm 40}$ & \\ 
    ATCA & 0.20 & 5.5 & $\bf{355 \pm 36}$ & \\ 
    ATCA & 0.24 & 5.5 & $\bf{367 \pm 37}$ & \\ 
    ATCA & 0.28 & 5.5 & $\bf{584 \pm 44}$ & \\ 
    ATCA & 0.33 & 5.5 & $\bf{500 \pm 48}$ & \\ 
    ATCA & 0.37 & 5.5 & $\bf{482 \pm 51}$ & \\ 
    ATCA & 0.41 & 5.5 & $\bf{337 \pm 88}$ & \\ 
    ATCA & 1.30 & 5.5 & $\bf{185 \pm 45}$ & 144 \\ 
    ATCA & 3.12 & 5.5 & $\bf{287 \pm 25}$ & 84 \\ 
    ATCA & 9.10 & 5.5 & $90 \pm 31$ & 144 \\ 
    ATCA & 16.08 & 5.5 & $16 \pm 19$ & 66 \\ 
    ATCA & 0.07 & 9.0 & $\bf{318 \pm 41}$ & \\ 
    ATCA & 0.11 & 9.0 & $\bf{388 \pm 36}$ & \\ 
    ATCA & 0.16 & 9.0 & $\bf{160 \pm 27}$ & \\ 
    ATCA & 0.20 & 9.0 & $\bf{139 \pm 26}$ & \\ 
    ATCA & 0.24 & 9.0 & $\bf{104 \pm 26}$ & \\ 
    ATCA & 0.28 & 9.0 & $\bf{171 \pm 26}$ & \\ 
    ATCA & 0.33 & 9.0 & $\bf{244 \pm 33}$ & \\ 
    ATCA & 0.37 & 9.0 & $\bf{463 \pm 40}$ & \\ 
    ATCA & 0.41 & 9.0 & $\bf{581 \pm 68}$ & \\ 
    ATCA & 1.30 & 9.0 & $\bf{212 \pm 32}$ & 99 \\ 
    ATCA & 3.12 & 9.0 & $\bf{235 \pm 24}$ & 81 \\ 
    ATCA & 9.10 & 9.0 & $70 \pm 26$ & 144 \\ 
    ATCA & 1.30 & 16.7 & $\bf{171 \pm 27}$ & 141 \\ 
    ATCA & 9.10 & 16.7 & $24 \pm 77$ & 420 \\
    \hline
    MeerKAT & 9.45 & 1.3 & $\bf{121 \pm 16}$ & 33 \\ 
    MeerKAT & 15.42 & 1.3 & $\bf{71 \pm 11}$ & 28 \\ 
    MeerKAT & 25.41 & 1.3 & $22 \pm 8$ & 27 \\ 
    MeerKAT & 40.39 & 1.3 & $17 \pm 8$ & 28 \\ 
    \hline
    AMI-LA & 0.60 & 15.5 & $128 \pm 62$ & 231 \\ 
    AMI-LA & 9.65 & 15.5 & & 295 \\ 
    \hline
    \textit{e}-MERLIN & 0.49 & 5.0 & $\bf{134 \pm 52}$ & 80 \\ 
    \textit{e}-MERLIN & 0.76 & 5.0 & $\bf{125 \pm 10}$ & 98 \\ 
    \textit{e}-MERLIN & 2.51 & 5.0 & & 120 \\ 
    \textit{e}-MERLIN & 6.09 & 5.0 & & 60 \\ 
    \textit{e}-MERLIN & 12.11 & 5.0 & $\bf{61 \pm 14}$ & 42 \\ 
    \textit{e}-MERLIN & 24.91 & 5.0 & & 62 \\ 
    \hline
    \end{tabular} 
    \caption{The radio density measurements of GRB 231117A. 
    The reported time post-burst corresponds to the logarithmic midpoint of the data point.
    The bolded flux densities indicate detections ($SNR>3$), and those not bolded are force-fitted flux density values ($SNR<3$). Note that force-fits could not be obtained from the \textit{e}-MERLIN data or the last AMI observation. 
    The error bars are $1\sigma$, and the $3\sigma$ threshold corresponds to 3 times the image RMS. 
    The first nine ATCA measurements at 5.5 and 9\,GHz were made in the $uv$-plane so we do not have a corresponding image RMS measurement. 
    }
    \label{tab:radio_flux}
\end{table*}

\begin{table}
    \centering
    \caption{Photometry of the host galaxy. All measurements are reported in the AB system and not corrected for reddening.}
    \begin{tabular}{cccc}
    \hline\hline
    Survey/   & Instrument    & Filter    & Brightness\\
    Telescope &             &           & (mag) \\
    \hline
    CFHT & MegaCAM & $u $ & $22.63 \pm 0.14 $\\
    PS1  &         & $g $ & $21.72 \pm 0.13 $\\
    PS1  &         & $r $ & $21.25 \pm 0.09 $\\
    PS1  &         & $i $ & $21.23 \pm 0.09 $\\
    PS1  &         & $z $ & $21.04 \pm 0.16 $\\
    PS1  &         & $y $ & $20.84 \pm 0.35 $\\
    LS   &         & $g $ & $21.75 \pm 0.04 $\\
    LS   &         & $r $ & $21.14 \pm 0.03 $\\
    LS   &         & $z $ & $21.03 \pm 0.06 $\\    
    VLT  & HAWK-I  & $K $ & $20.68 \pm 0.10 $\\    
    \hline\hline
    \end{tabular}
    \label{tab:phot:host}
\end{table}

\section{Violent collision model}\label{ap:model}

\begin{figure}
    \centering
    \includegraphics[width=\linewidth]{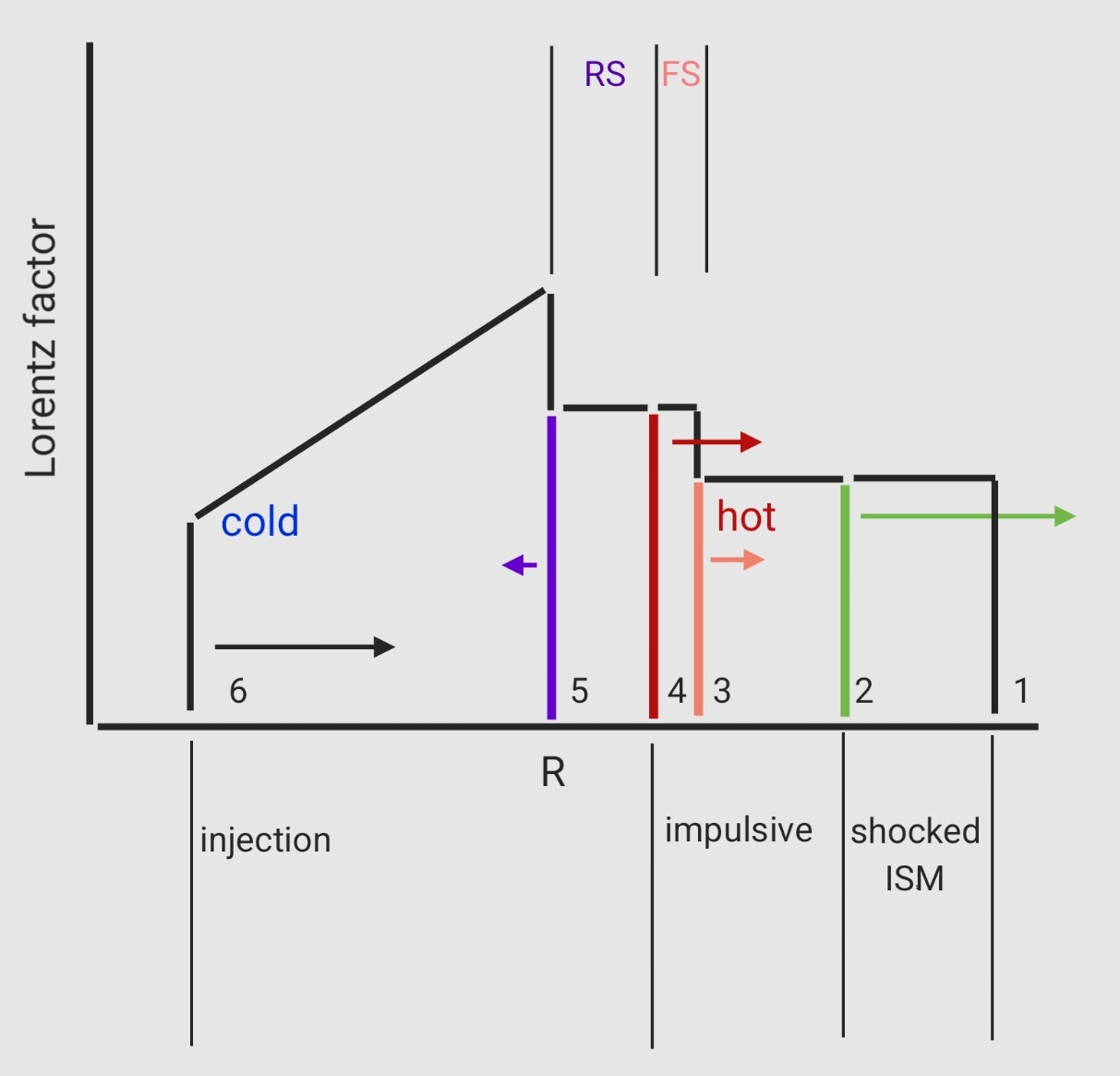}
    \caption{A schematic diagram of the colliding relativistic shells. The numbers indicate the zones: 1 -- unshocked ISM, 2 -- shocked ISM, 3 -- hot impulsive shell, 4 -- shocked hot impulsive shell, 5 -- shocked injection shell, 6 -- unshocked injection shell. The coloured lines indicate:
    impulsive shell -- shocked ISM contact discontinuity (green); 
    hot impulsive shell shocked -- hot contact discontinuity (orange);
    impulsive shell -- injection contact (red);
    reverse shock -- unshocked injection shell contact discontinuity (purple).
    RS and FS indicate the forward- and reverse-shock system due to the violent collision, where the forward shock is into hot material while the reverse shock is through a cold shell. The axes are not to scale.}
    \label{fig:collision_schem}
\end{figure}

The case where energy injection results in a violent collision (i.e., new strong shocks are generated within both an impulsive (caught) and injection (catching) shell of relativistic and jetted outflow) is well derived analytically in \cite{kumar2000} and \cite{zhang2002} \citep[see also,][for long lived reverse shocks in a stratified ejecta shell]{uhm12}.
Here we adopt the notation used in \cite{zhang2002} for our shocked system, see Figure\,\ref{fig:collision_schem} for zone, shell, and shock identifiers -- shown as a sketch of shell Lorentz factor with radius.

For our collision scenario, we assume that the shells were launched at the same time, but due to differences in the Lorentz factor become radially separated.
As the initial and impulsive shell decelerates, the slower injection shell will catch-up when $\gamma_3 = \gamma_{6,0}/2$.
For a thin shell, the deceleration radius is $R_{\rm dec} \propto \Gamma_0^{-2/3}$, and the time scale is then, $t_{\rm dec} \sim R_{\rm dec} / ( 2\Gamma_0^2 c)$ -- here $\Gamma_0$ is the initial Lorentz factor of the impulsive shell.
The collision radius is then $R_{\rm col} \sim R_{\rm dec} (\gamma_{\rm collison}/\Gamma_0)^{-2/3}$, where $\gamma_{\rm collison}$ is an input from the refreshed shock model fit.
The collision time is then \citep{zhang2002},
\begin{equation}
    \Delta T_{36} = R_{\rm col} \frac{1 + \gamma_{3,0}^2}{\gamma_{6,0}^2} \frac{1}{2 \gamma_{3,0}^2 c},
\end{equation}
where $\gamma_{3,0}$ and $\gamma_{6,0}$ are the Lorentz factors of the impulsive and the catching front of the injection shell at collision.

The timescale over which energy injection occurs is found from the final Lorentz factor of the injection shell on collision is 
\begin{equation}
    \gamma_{6,{\rm end}} = 1 + (\gamma_{6,0} - 1) \zeta^{1/(1-s)},
\end{equation}
where $\zeta$ is the factor by which energy increases due to injection, and $s$ is the index for the increase in mass.
The radius for the end of injection is then,
\begin{equation}
    R_{\rm inj. end} = R_{\rm col} \left(\frac{\gamma_{6,{\rm end}}}{\gamma_{6,0}}\right)^{-(s+1)/3},
\end{equation}
and the corresponding observer time,
\begin{equation}
    t_{\rm inj. end} = (t_{\rm dec} + \Delta T_{36}) \left(\frac{\gamma_{6,{\rm end}}}{\gamma_{6,0}}\right)^{-(s+7)/3},
\end{equation}
and the injection timescale, $\Delta t_{\rm inj} = t_{\rm inj. end} - \Delta T_{36}$.
The injection shell thickness is then $\Delta_6 \sim \Delta t_{\rm inj} c$.

The swept-up ISM mass, the shocked ISM, is always $M_2 \propto R^3$.
The unshocked impulsive shell will continue to decelerate as $\gamma_3 \propto R^{-3/2}$, and the unshocked injection shell Lorentz factor evolves as $\gamma_6 \propto R^{-3/(1+s)}$, while injection occurs $(t_{\rm col} < t < t_{\rm inj. end})$.
The mass of the reverse shocked injection shell, during the violent collision stage, is given by $M_5 \propto \gamma_6^{-s}$.

The Lorentz factors and the various regions of our violent collision system are shown schematically in Figure\,\ref{fig:collision_schem}. Here the numbers refer to: the unshocked ISM (1); the shocked ISM (2); the hot `unshocked' impulsive shell (3); the forward shocked, hot impulsive shell (4); the reverse shocked injection shell (5); and the unshocked injection shell (6). The condition for violent injection is $e_4/e_3 > 1$ and $\gamma_{36} > \sqrt{3/2} \sim 1.225$, the supersonic shock condition for a relativistic equation of state \citep[see Fig. 2 in ][]{zhang2002} -- during energy injection, we find the masses and the Lorentz factors for each shocked region while both of these conditions are met, and assume `mild' energy injection when not.

The ratio of the energy densities of region 3 and 4 is found by solving equations \ref{eq:eratio} and \ref{eq:gamma34} giving the instantaneous values of $e_4/e_3$, $\gamma_4$, and $\gamma_{43} = (\gamma_4/\gamma_3 + \gamma_3/\gamma_4)/2$ (the Lorentz factor of the hot, forward shocked shell (4) as seen co-moving with (3), the impulsive shell).

While $e_4/e_3 > 1$ then the mass in region 6 is given by $M_6 = M_{6,0} - M_5$, and the shocked mass in region 4 is found by considering the energy conservation of the entire system,
\begin{equation}
    M_4 = \frac{M_5\left(\gamma_{6,0} - \gamma_{56}\gamma_4\right) + M_{3,0} \left(\gamma_{3,0} - \gamma_3\right) - M_2 \left(\gamma_3^2 - 1\right)}{\gamma_3\left(\gamma_4 - 1\right)},
    \label{eq:conserve}
\end{equation}
where the denominator $\gamma_4\gamma_3 - \gamma_3$ accounts for the energy of the shocked hot material within the impulsive shell and ensures mass conservation throughout the violent collision stage.

When $M_4 = M_{3,0}$, the forward shock has crossed the impulsive shell and region 3 is replaced with region 4.
If the crossing radius for the forward shock is $R_{\rm x} < R_{\rm inj. end}$, then injection continues, however, the new shock conditions need to be solved between region 4 and 6.
As $\gamma_{56} \lesssim 1.225$ and $\gamma_4 = \gamma_5$, then the supersonic condition is no longer met and energy injection follows the mild case.
From this point any energy injected into the impulsive shell is added directly to region 4, this increases the $M_4$ as injection continues.
The hot impulsive shell will now decelerate as $\gamma_4 \propto R^{-3/2}$, however, as $\gamma_4 > \gamma_3$, the additional mass swept up by the external forward shock, $M_2$, will have a higher relativistic mass $\equiv \gamma M_2$.
This is the same as the scenario modelled via a refreshed shock with energy injection as $E \propto \gamma^{1-s}$.

The earlier reverse shocked matter is no longer sustained, and so the hot matter in region 5 remains constant, $M_5 = M_5(R_{\rm x})$, but the region cools as $\gamma_5 \propto R^{-g}$, where $3/2 < g < 7/2$, and emission from (5) will continue until the shocked electrons are efficiently cooled \citep{kobayashi00sari}.
The impulsive region continues to gain energy as mass is added from the injection shell, until injection ends.
The external shock system is then governed by the deceleration of a shell with mass $= M_{6,0} + M_{3,0}$ and $\gamma_4 \propto R^{-3/2}$.
The Lorentz factor of the earlier reverse shock will then evolve as $\gamma_{56} \propto R^{-3/2}$.

\subsection{Emission during the violent collision}
There are two emission sites that we consider during the violent collision episode.
The non-thermal emission from the reverse shock region 5 and the hot forward shock region 4.
\subsubsection{Reverse shock}
For the emission from the reverse shock, the total number of shocked electrons (assuming charge neutrality) is $N_{e,5} = M_5/m_p$, where $m_p$ is the mass of a proton.
The number density in the injection shell is $n_6 \sim M_6/(4\pi R^2\Delta_6 \gamma_6 m_p)$ and then, following the shock jump conditions, $n_5 = n_6(4\gamma_{56} + 3)$ -- where we have assumed an adiabatic index $\hat\gamma = 4/3$.
When the violent collision condition ends, then the ambient density of the reverse region evolves as $n_5 \propto R^{-3-g}$ until the end of energy injection, when $n_5 \propto R^{-9/2}$.

The emission from the reverse shock is calculated using the standard synchrotron emission relations for relativistic plasma \citep[e.g.,][]{sari98}.
We additionally use the reverse shock coefficients given by \cite{harrison2013};
these rely on the dimensionless parameter, $\varepsilon = (l/\Delta)^{1/2} \gamma^{-4/3}$, where $l = (3 E_{\rm total} / [4\pi m_p n c^2 ] )^{1/3}$ is the Sedov length.
The coefficients are,
\begin{eqnarray}
    &C_{\rm f} &= 1/\left({3/2 + 5\varepsilon^{-1.3}}\right), \\
    &C_{\rm m} &= 0.005 + \varepsilon^{-3}, \\
    &C_{\rm t} &= 0.2 + \varepsilon^{-2},
    \label{eq:harrison}
\end{eqnarray}
where these are used to multiply the maximum flux, $C_{\rm f}$, the minimum synchrotron frequency, $C_{\rm m}$, and the observer timescale for the reverse shock, $C_{\rm t}$.

The magnetic field for the reverse shock region is,
\begin{equation}
    B_5^2 = {8\pi m_p R_B \varepsilon_B n_5 c^2 \frac{\hat\gamma \gamma_{56} +1}{\hat\gamma - 1} \left(\gamma_{56} -1 \right)}.
\end{equation}
The power per electron is,
\begin{equation}
    P_5 = \frac{f_x m_e c^2 \sigma_T B_5}{3 q},
\end{equation}
where $f_x$ is the dimensionless peak flux from \cite{Wijers99}, $m_e$ is the mass of an electron, $\sigma_T$ the Thomson cross-section, and $q$ the charge of an electron.
The maximum synchrotron flux is then,
\begin{equation}
    F_{\rm RS,max} = \frac{\Xi_{N,5} C_{\rm f} N_{e,5} P_5 \gamma_5^2}{\pi D_L^2},
\end{equation}
here, $D_L$ is the luminosity distance.
As we consider a broad prior on the accelerated electron distribution index, $1.3 < p < 3.4$, then to calculate the characteristic frequency we need to consider the conditional factors \citep[e.g.,][]{gao13},
\begin{equation}
    g_m = \left(\frac{6 \pi q}{\sigma_T B_5}\right)^{1/2}, 
\end{equation}
\begin{eqnarray}
    g_p =& 
\begin{dcases}
    \frac{(p-2)}{(p-1)}. ~~~&p > 2,\\
    \left[\ln\left(\frac{\Xi_{N,5}~ g_m}{\varepsilon_e\left(\gamma_{56} - 1\right)}\frac{ m_e}{m_p}\right)\right]^{-1},  ~~~&p = 2,\\
    \frac{(2 - p)}{(p -1)}\frac{m_p}{m_e} g_m^{p - 2}, ~~~&p < 2,\\
\end{dcases}\nonumber
\\
   \gamma_{m,5} =& 
\begin{dcases}
    \frac{\varepsilon_e}{\Xi_{N,5}}\frac{m_p}{m_e} ~g_p \left(\gamma_{56} - 1\right),  ~~~~~&p \geq 2,\\
    \left[\frac{g_p ~\varepsilon_e (\gamma_{56} - 1)}{\Xi_{N,5}} \right]^{1/(p-1)},   ~~~~~&p < 2.
\end{dcases}
\nonumber
\end{eqnarray}

The synchrotron characteristic frequency is then,
\begin{equation}
    \nu_{m,5} = \frac{3 x_p C_{\rm m} \gamma_5 q B_5 \gamma_{m,5}^2}{2 \pi m_e c},
\end{equation}
here, $x_p$ is the dimensionless maximum of the spectrum \citep{Wijers99}.
Once the violent collision conditions are no longer met, then $\nu_{m,5} \propto t^{-73/48}$ and $F_{\rm RS,max} \propto t^{-47/48}$, where $t$ is the laboratory time -- $t = t_{\rm obs}/(1 + z)$.
Emission at frequencies $\nu > \nu_c$ are cut-off, as the electrons are efficiently cooled, and there is no new supply of electrons within the shock.
Before the shock ends, the cooling frequency is,
\begin{eqnarray}
    \gamma_c =& {6\pi m_e c}/({\sigma_T \gamma_{56} B_5^2 t}),&\\
    \nu_c =& 3 (0.286) \gamma_c^2 {\gamma_5 q B_5}/({2 \pi m_e c}),&
\end{eqnarray}
where we follow \cite{Wijers99} for the derrivation of the cooling frequency. 

Synchrotron self-absorption is approximated by considering the blackbody flux \citep{lambkobayashi19},
\begin{equation}
    F_{\rm RS,BB} = 4\pi~\gamma_5 \gamma_{m,5} m_e ~\nu^2~\frac{R^2}{D_L^2} \text{max}[1,(\nu/\nu_{m,5})^{1/2}],
\end{equation}
where $\nu$ is the emission frequency.
The flux density from the reverse shock is then the minimum of $F_{{\rm RS,BB}}(\nu)$ and $F_{\nu}$ at each band.

\subsubsection{Forward shock}
Within the impulsive shell, emission from the forward shock only becomes significant some time after the collision.
As the impulsive shell is thin, once the forward shock is established, it rapidly crosses the shell -- after which the violent collision conditions are no longer met for our parameters and the shock system fails.

Within the impulsive shell, which is thin, the number density is $n_3 = 4 n_1 \gamma_{3,0}^2 (t_{\rm F}/t_{\rm col})^{-3}$, where $t_{\rm F}$ is the laboratory time for the forward shock emission \citep{kobayashi2000lightcurves}.
The shocked particle density is $n_4 = n_3 ([e_4/e_3 (1 + e_4/e_3)]/[3 + e_4/_3])^{1/2}$, and the emission is governed by the random Lorentz factor defined as $\gamma_B = e_4/(n_4m_pc^2)$ \citep{zhang2002} and given by,
\begin{equation}
    \gamma_B \sim \gamma_{3,s}M_4 {(\gamma_{43} - 1)}/({4\pi R^2 \gamma_4^2 \Delta_{3} n_4 m_p}),\\
    \label{eq:gammaB}
\end{equation}
here, $\gamma_{3,s}$ is the impulsive shell Lorentz factor when the forward shock emission is established (i.e., $M_4 \gg 0$, and becomes significant).
The magnetic field for the forward shock is,
\begin{equation}
    B_4^2 = 8\pi m_p \varepsilon_B n_4 c^2 \frac{\hat\gamma \gamma_B +1}{\hat\gamma - 1} (\gamma_B - 1).
\end{equation}
The synchrotron power per electron is $P_4 = f_x m_e c^2 \sigma_T B_4/(3 q)$ and the maximum flux,
\begin{equation}
    F_{\rm FS,max} = \frac{\Xi_{N,4} N_{e,4} \gamma_4^2 P_4}{\pi D_L^2}.
\end{equation}

To ensure the forward shock emission is consistent with the observed flux density levels, we force $\Xi_{N,4} = 0.1 \Xi_N$.
The flux density is then calculated using the same approximations as for the reverse shock but replacing $\gamma_{56} \rightarrow \gamma_B$.
Similarly, for the blackbody flux $\gamma_{m,5} \rightarrow \gamma_{m,4}$.
For the forward shock, once the forward shock has crossed the impulsive shell, the shock system dies and the random Lorentz factor, $\gamma_B \sim \gamma_4$ and the shell evolves adiabatically with the mild energy injection.
The forward shock in our scenario results in a flare as the shock crosses the impulsive shell.
Synchrotron self-absorption is considered in the same way as for the reverse shock.


\bibliography{papers}{}
\bibliographystyle{aasjournal}



\end{document}